%% file: 1main_doc.tex
\documentclass[twocolumn]{aastex62}

\received{March 6, 2018}
\revised{June 8, 2018}
\accepted{June 9, 2018} %
\submitjournal{ApJ}

\shorttitle{Galaxy and AGN Luminosity Functions at $z=4$ with SHELA}
\shortauthors{Stevans et al.}

\begin{document}

\title{Bridging Star-Forming Galaxy and AGN Ultraviolet Luminosity Functions at $z=4$\\ with the SHELA Wide-Field Survey}

 \correspondingauthor{Matthew L. Stevans}
 \email{stevans@utexas.edu}

 \author[0000-0001-8379-7606]{Matthew L. Stevans}
 \affiliation{Department of Astronomy, University of Texas at Austin, Austin, TX 78705}
 
  \author[0000-0001-8519-1130]{Steven L. Finkelstein}
 \affiliation{Department of Astronomy, University of Texas at Austin, Austin, TX 78705}

\author{Isak Wold} 
 \affiliation{Department of Astronomy, University of Texas at Austin, Austin, TX 78705}
 
\author[0000-0003-4032-2445]{Lalitwadee Kawinwanichakij}
\affiliation{Department of Physics and Astronomy, Texas A\&M University, College
Station, TX, 77843-4242 USA}
\affiliation{George P.\ and Cynthia Woods Mitchell Institute for
  Fundamental Physics and Astronomy, Texas A\&M University, College
  Station, TX, 77843-4242 USA}
\affiliation{LSSTC Data Science Fellow}  

\author[0000-0001-7503-8482]{Casey Papovich}
\affiliation{Department of Physics and Astronomy, Texas A\&M University, College
Station, TX, 77843-4242 USA}
\affiliation{George P.\ and Cynthia Woods Mitchell Institute for
  Fundamental Physics and Astronomy, Texas A\&M University, College
  Station, TX, 77843-4242 USA}

\author{Sydney Sherman}
 \affiliation{Department of Astronomy, University of Texas at Austin, Austin, TX 78705}
 
\author{Robin Ciardullo}
 \affiliation{Department of Astronomy \& Astrophysics, The Pennsylvania State University, University Park, PA 16802}
 \affiliation{Institute for Gravitation and the Cosmos, The Pennsylvania State University, University Park, PA 16802}

\author{Jonathan Florez}
 \affiliation{Department of Astronomy, University of Texas at Austin, Austin, TX 78705}
 
\author{Caryl Gronwall}
 \affiliation{Department of Astronomy \& Astrophysics, The Pennsylvania State University, University Park, PA 16802}
 \affiliation{Institute for Gravitation and the Cosmos, The Pennsylvania State University, University Park, PA 16802}

\author{Shardha Jogee}
 \affiliation{Department of Astronomy, University of Texas at Austin, Austin, TX 78705}
 
\author{Rachel S. Somerville}
 \affiliation{Department of Physics \& Astronomy, Rutgers University, Piscataway, New Jersey 08854}
 \affiliation{Center for Computational Astrophysics, Flatiron Institute, New York, NY, 10010}
 
\author{L. Y. Aaron Yung}
 \affiliation{Department of Physics \& Astronomy, Rutgers University, Piscataway, New Jersey 08854}

\input{s0_abstract.tex}


\input{s1_intro.tex}

\input{s2_reduction.tex}

\input{s3_sample.tex}

\input{s4_results.tex}

\input{s5_discussion.tex}

\input{s6_conclusions.tex}

\acknowledgments
The authors acknowledge Raquel Martinez, Karl Gebhardt, and Eric J. Gawiser for the interesting discussions we had and their suggestions which improved the quality of this research. We thank Neal J. Evans and William P. Bowman for useful comments on the draft of this paper. M. L. S. and S. L. F. acknowledge support from the University of Texas at Austin, the NASA Astrophysics and Data Analysis Program through grant NNX16AN46G, and the National Science Foundation AAG Award AST-1614798. The work of C. P. And L. K.  is supported by the National Science Foundation through grants AST 413317, and 1614668. The Institute for Gravitation and the Cosmos is supported by the Eberly College of Science and the Office of the Senior Vice President for Research at the Pennsylvania State University. S. J., S. S. and J. F. acknowledge support from the University of Texas at Austin and NSF grants NSF AST-1614798 and NSF AST-1413652. R. S. S. and A. Y. thank the Downsbrough family for their generous support, and gratefully acknowledge funding from the Simons Foundation.

\vspace{5mm}
\facilities{CTIO,SST(IRAC)}

\bibliographystyle{aasjournal1}

\bibliography{1main-doc.bib}

\end{document}

%% file: s0_abstract.tex
\begin{abstract}
We present a joint analysis of the rest-frame ultraviolet (UV) luminosity functions of continuum-selected star-forming galaxies and galaxies dominated by active galactic nuclei (AGNs) at $z \sim$ 4.  These 3,740 $z \sim$ 4 galaxies are selected from broad-band imaging in nine photometric bands over 18 deg$^2$ in the {\it Spitzer}/HETDEX Exploratory Large Area Survey (SHELA) field.  The large area and moderate depth of our survey provide a unique view of the intersection between the bright end of the galaxy UV luminosity function (M$_{AB}<-$22) and the faint end of the AGN UV luminosity function.  We do not separate AGN-dominated galaxies from star-formation-dominated galaxies, but rather fit both luminosity functions simultaneously.  These functions are best fit with a double power-law (DPL) for both the galaxy and AGN components, where the galaxy bright-end slope has a power-law index of $-3.80\pm0.10$, and the corresponding AGN faint-end slope is $\alpha_{AGN} = -1.49^{+0.30}_{-0.21}$. We cannot rule out a Schechter-like exponential decline for the galaxy UV luminosity function, and in this scenario the AGN luminosity function has a steeper faint-end slope of $-2.08^{+0.18}_{-0.11}$. Comparison of our galaxy luminosity function results with a representative cosmological model of galaxy formation suggests that the molecular gas depletion time must be shorter, implying that star formation is more efficient in bright galaxies at $z=4$ than at the present day. If the galaxy luminosity function does indeed have a power-law shape at the bright end, the implied ionizing emissivity from AGNs is not inconsistent with previous observations. However, if the underlying galaxy distribution is Schechter, it implies a significantly higher ionizing emissivity from AGNs at this epoch. 
\end{abstract}

%% file: s1_intro.tex
\section{Introduction} \label{intro}

Explaining how galaxies grow and evolve over cosmic time is one 
of the main goals of extragalactic astronomy. With the number of massive galaxies increasing from $z\sim4-2$ \citep{marchesini09,muzzin13} and the positive relation between stellar mass and star formation rate, by studying the properties of galaxies with the highest star formation rates at $z\sim4$ we can glean how the most massive galaxies built up their stellar mass. The use of multi-wavelength photometry and the Lyman break technique has revolutionized the study of galaxies in the $z>2$ universe
\citep[e.g.,][]{steidel96}. These tools are currently
the most efficient for selecting large samples of high-redshift star-forming galaxies for extensive study. A power tool for understanding the distribution of star-formation at high redshifts is the rest-frame UV luminosity function. This probes recent unobscured star-formation directly over the last 100 Myr and is, therefore, a fundamental tracer of galaxy evolution.

The shape of the star-forming galaxy UV luminosity function
at $z=4$ has been difficult to pin down at the bright end. The characteristic luminosity of the Schechter function, which is often used to describe the luminosity function in field environments, ranges over a few orders of magnitude \citep[e.g.,][]{steidel99,fink15,viironen17}, and there is growing evidence of an excess of galaxies over the exponentially declining bright end of the Schechter function \citep[e.g.,][]{van10,ono17}. The uncertainty at the bright end is due in part to cosmic variance and the small area of past surveys which miss the brightest galaxies 
with the lowest surface density. The largest 
$z\sim4$ spectroscopically observed sample used in a published luminosity function is from the VIMOS VLT Deep Survey (VVDS, \citealt{lefevre13}) consisting of 129 spectra from $\sim1$ deg$^{2}$ \citep{cucciati12}. This small 
sample size limits the analysis of how galaxy growth 
properties (e.g., star-formation rate (SFR)) depend on properties like stellar 
mass and environment, especially at the bright-end. 
The large cost of spectroscopically surveying faint 
sources leaves the most efficient method of using 
multi-wavelength photometry as the best way to collect 
larger samples of star-forming galaxies. For example, a few thousand $z=4$ galaxies were 
detected in the four 1 deg$^2$ fields of the CFHT 
Survey \citep{van10}.

Another challenge in measuring the bright end of the UV luminosity function is the existence of AGNs and their photometric similarities with UV-bright galaxies. The spectral energy distributions (SEDs) of AGN-dominated galaxies are characterized by a power-law continuum and highly ionized emission lines in the rest-frame UV \citep[e.g.,][]{stevans:2014}, and like high-$z$ UV-bright galaxies the observed SEDs of high-$z$ AGNs exhibit a Lyman break feature due to absorption from intervening neutral hydrogen in the IGM. Thus, any UV-bright galaxy selection technique relying on the Lyman break will also select AGNs. Some have attempted to use a morphological cut to break the color degeneracy of UV-bright galaxies and AGNs by assuming the former will appear extended and the latter will be strictly point sources. However, this method is less reliable near the photometric limit especially in ground-based imaging with poor seeing. For example, recent work by \citet{akiyama17} has shown such a morphological selection can select a sample of point sources with only 40\% completeness and 30\% contamination at $i=24$ mag in photometry with median seeing conditions of 0\farcs7 and  5-$\sigma$ depths of $i=26.4$ mags. 

The shape of the AGN luminosity function is of interest as well, as a steep faint end can result in a non-negligible contribution of ionizing photons from AGNs to the total ionizing budget.  Current uncertainties in the literature at $z \geq$ 4 are large (\citealt{glikman11}, \citealt{masters12}, \citealt{giallongo15}), thus AGNs have received renewed interest in the literature with regards to reionization \citep[e.g.,][]{madau15,mcgreer17}.

Studying both AGN-dominated and star-formation-dominated UV-emitting galaxies simultaneously is possible given a large enough volume. The Hyper Suprime-Cam (HSC) Subaru strategic program (SSP) has detected large numbers of both types of objects at $z=4$ using their optical-only data.  However, \citet{akiyama17} opt to use their excellent ground-based resolution (0.6$\arcsec$; 4.27 kpc at $z=4$) to remove extended sources and focus on the AGN population separately. \citet{ono17} selects $z\sim4$ galaxies (and AGNs) as $g'$-band dropouts in the HSC SSP using strict color cuts including the requirement that sources are not significantly detected ($\sigma<2$) in the $g'$ band. This $g'$ band could remove UV-bright AGNs and could explain why \citet{ono17} find less sources at $M<-24$ mag than \citet{akiyama17} (see Figure 7 in \citealt{ono17}).

Here, we make use of the 24 deg$^2$ \textit{Spitzer}-HETDEX Exploratory Large Area (SHELA) survey dataset to probe both AGN-dominated and star-formation-dominated UV-emitting galaxies over a large area. The SHELA dataset includes deep (22.6 AB mag, 50\% complete) 3.6 $\mu$m and 4.5 $\mu$m  imaging from \textit{Spitzer}/IRAC \citep{papovich16} and $u^{\prime}g^{\prime}r^{\prime}i^{\prime}z^{\prime}$ imaging from the Dark Energy Camera over 18 deg$^2$ (DECam; Wold et al.\ , in prep). Because SHELA falls within SDSS Stripe-82 there exists a large library of ancillary data, which we take advantage of by including in our analysis the VISTA J and Ks photometry from the VICS82 survey \citep{geach17} to help rule out low-$z$ interloping galaxies. In addition, there is deep X-ray imaging in this field from the Stripe-82X survey \citep{lamassa16}, which could be used to identify bright AGNs.

We select objects at $z >$ 4 based on photometric criteria. Our sample includes, therefore, both galaxies whose light is powered by star-formation and AGN activity. As the bulk of AGNs at $z\sim4$ are too faint to detect in the existing X-ray data, we include all $z=4$ candidate galaxies, regardless of powering source, in our sample and use our large dynamic range in luminosity -- combined with very bright AGNs from SDSS (\citealt{richards06}, \citealt{akiyama17}) and very faint galaxies from the deeper, narrower {\it Hubble Space Telescope} surveys (\citealt{fink15}) -- to fit the luminosity functions of both populations simultaneously.  Importantly, our sample is selected using both optical and {\it Spitzer} mid-infrared data, which results in an improved contamination rate over optical data alone.

This paper is organized as follows. The SHELA field 
dataset used in this paper is summarized in \ref{dataset}.
The DECam reduction are 
discussed in Sections \ref{data_reduction}--\ref{phot_errors} and the IRAC data reduction and photometry in \ref{irac}. Sample selection and contamination are 
discussed in Section \ref{sample}. Our UV luminosity 
function is presented in Section \ref{results}. The implications of our results are discussed in section \ref{discussion}. We 
summarize our work and discuss future work in Section 
\ref{conclusions}. Throughout this paper we assume a Planck 
2013 cosmology, with H$_0 = 67.8$ km s$^{-1}$ Mpc$^{-1}$,
$\Omega_M = 0.307$ and $\Omega_{\Lambda} = 0.693$ 
\citep{planck13}. All magnitudes given are in the AB 
system \citep{oke83}.

%% file: s2_reduction.tex
\section{Data Reduction and Photometry} \label{data}

In this section, we describe our dataset, image reduction, and source extraction procedures. The procedures applied to DECam imaging are largely similar to those used with these data in Wold et al.\ (in prep), thus we direct the reader there for more detailed information.

\subsection{Overview of Dataset} \label{dataset}
In this study, we use imaging in nine photometric bands spanning the optical to mid-IR in the SHELA Field. The SHELA Field  is centered at R.A. = $1^{\textrm{h}}22^{\textrm{m}}00^{\textrm{s}}$, declination = +0\arcdeg00\arcmin00\arcsec (J2000) and extends approximately $\pm6\fdg5$ in R.A. and $\pm1\fdg25$ in declination. The optical bands consist of $u'$, $g'$, $r'$, $i'$, and $z'$ from the Dark Energy Camera (DECam) and covers $\sim$17.8 deg$^2$ of the SHELA footprint (Wold et al.\ in prep). The mid-IR bands include the 3.6 $\mu$m and 4.5 $\mu$m from the Infrared Array Camera (IRAC) aboard the \textit{Spitzer Space Telescope} and covers 24 deg$^2$ \citep{papovich16}. In addition, we include near-IR photometry in $J$ and $K_{s}$ from the February 2017 version of the VISTA-CFHT Stripe 82 Near-infrared Survey\footnote{http://stri-cluster.herts.ac.uk/vics82/} (VICS82; \citealt{geach17}), which covers $\sim$85\% of the optical imaging footprint and has 5-$\sigma$ depths of $J=21.3$ mag and $K_{s}=20.9$ mag. Figure \ref{filters} shows the filter transmission curves for the nine photometric bands used overplotted with model high-$z$ galaxy spectra illustrating the wavelength coverage of our dataset.

\input{fig_filters.tex}

\subsection{DECam Data Reduction and Photometric Calibration} \label{data_reduction}
The DECam images were processed by the NOAO Community Pipeline (CP). A detailed description of the Community Pipeline reduction procedure can be found in the DECam Data Handbook on the NOAO website\footnote{http://ast.noao.edu/data/docs}, however, we provide a brief summary of the procedure here.  First, the DECam images were calibrated using calibration exposures from the observing run. The main calibration steps included an electronic bias calibration, saturation masking, bad pixel masking and interpolation, dark count calibration, linearity correction, and flat-field calibration. Next, the images were astrometrically calibrated with 2MASS reference images. Finally, the images were remapped to a grid where each pixel is a square with a side length of 0.27$\arcsec$. Observations taken on the same night were then co-added.

\input{table_depth_fwhm.tex}

The CP data products for the SHELA field were downloaded from the NOAO Science Archive\footnote{http://archive.noao.edu/}. The data products include the co-added images, remapped images, data quality maps (DQMs), exposure time maps (ETMs), and weight maps (WMs). The co-added images from the CP were not intended for scientific use, so we opted to co-add the remapped images. We followed the co-adding procedure of Wold et al. (in prep.), which we summarize here. Using the software package SWARP \citep{bertin02} the sub-images stored in the FITS files of the remapped images were stitched together and background subtracted. The remapped images were combined using a weighted mean procedure optimized for point-sources. The weighting of each image is a function of the seeing, transparency, and sky brightness and is defined by Equation A3 in \citet{gawiser06} as
\begin{equation}
w^{\mathrm{PS}}_{i} = \Bigg( \frac{\mathrm{factor}_i}{\mathrm{scale}_i \times \mathrm{rms}_i}\Bigg)^2,
\label{weightps}
\end{equation}

\noindent where $\mathrm{scale}_i$ is the image transparency (defined as the median brightness of the bright unsaturated stars after normalizing the brightness measurement of each star by its median brightness across all exposures), $\mathrm{rms}_i$ is the root mean square of the fluctuations in background pixels, and $\mathrm{factor}_i$ is defined as

\begin{equation}
\mathrm{factor}_i = 1 - \mathrm{exp} \Bigg( -1.3 \frac{\mathrm{FWHM}^{2}_{\mathrm{stack}}}{\mathrm{FWHM}^{2}_{i}} \Bigg),
\end{equation}

\noindent where $\mathrm{FWHM}_{\mathrm{stack}}$ is the median FWMH of bright unsaturated stars in an unweighted stacked image and $\mathrm{FWHM}_{i}$ is the median FWHM of bright unsaturated stars in each individual exposure.

The seeing and transparency measurements were determined using a preliminary source catalog generated for each resampled image using the Source Extractor software package \citep{bertin96}.The seeing in the final stacked images are listed in Table \ref{depth_table}. 

After discovering the original WMs from the Community Pipeline had values inconsistent with the ETM, we created custom rms maps for the co-added images. The initial rms per pixel was defined as the inverse of the square-root of the exposure time. The median of the rms map is scaled to the global pixel-to-pixel rms which is defined as the standard deviation of the fluxes in good-quality, blank sky pixels. Good-quality, blank sky pixels are pixels not included in a source according to our initial Source Extractor catalog (see Section \ref{sextration} for discussion of our source extraction procedure), and have an exposure time greater than 0.9 times the median value.

The DECam imaging data were photometrically calibrated with photometry from the Sloan Digital Sky Survey (SDSS) data release 11 (DR11; \citealp{eisenstein11}) using only F0 stars. F0 stars were used because their spectral energy distribution span all five optical filters while appearing in the sky at a sufficiently high surface density to provide statistically significant numbers in each DECam image. We began by creating a preliminary source catalog for the stacked DECam images using Source Extractor and position matching to the SDSS source catalog. Then we selected F0 stars using SDSS colors by integrating an F0-star model spectrum from the 1993 Kurucz Stellar Atmospheres Atlas \citep{Kurucz79} with each of the five optical SDSS filter curves. For sources in the catalog to be identified as an F0 star, the total color differences, using colors for all adjacent bands, added in quadrature must have been less than 0.35. We then calculate the expected magnitude offset between SDSS and DECam filters for F0 stars, which are as follows: $u^{\prime}$: 0.33, $g^{\prime}$: 0.02, $r^{\prime}$: -0.001, $i^{\prime}$: -0.02, and $z^{\prime}$: -0.01. The zero-point for each filter was then calculated as
\begin{equation}
\mathrm{ZPT} = \mathrm{median}(m^{AB}_{\mathrm{SDSS}}-m_{\mathrm{DECam}}-\Delta m_{\mathrm{offset}}),
\label{zpt}
\end{equation}
\noindent where $\Delta m_{\mathrm{offset}}$ is the expected magnitude offset between SDSS and DECam filters for F0 stars. After the zero-points were applied to each stacked image, the image pixel values were converted to units of nJy.

\input{fig_psf-compare.tex}

\subsection{DECam Photometry} \label{sextration} 

Studying galaxy properties relies on accurate measurements of galaxy colors. One way to obtain accurate colors is to perform fixed-aperture photometry where you measure source fluxes in every band using the same sized aperture. However, since the DECam images where taken in varying seeing conditions they have point spread functions (PSFs) with a range of full width at half maximum (FWHM). To perform fixed-aperture photometry on these images, the PSFs of all the imaging covering a single patch of sky must be adjusted to have a similar PSF size. We divided the DECam imaging into six sub-fields (each defined as roughly one DECam pointing). For each sub-field, we enlarged the PSFs of the stacked images to match the PSF of the stacked image with the largest PSF in that sub-field. For example, in sub-field SHELA-1 we matched the PSFs of the $g^{\prime}$, $r^{\prime}$, $i^{\prime}$, and $z^{\prime}$ stacked images to the PSF of the  $u^{\prime}$ band. To enlarge the PSFs we adopted the procedure of \citet{fink15} who used the IDL deconv\_tool Lucy-Richardson deconvolution routine. This routine takes as inputs two PSFs (the desired larger PSF and the starting smaller PSF) and the number of iterations to run and outputs a convolution kernel. The input image PSFs were produced by median combining the 100 brightest stars (sources with stellar classifications in SDSS DR11) in each image. Before combining, the stars were over-sampled by a factor of 10, re-centered, and then binned by ten to ensure the star centroids aligned. We ran the deconvolution routine with an increasing number of iterations until the PSF of the convolved image (again measured from stacking stars) had a flux within a 7-pixel (1.89\arcsec) diameter aperture matched to that of the PSF of the target image to within 5\%. The total fluxes were measured in 30-pixel (8.1\arcsec) diameter circular apertures. In Figure \ref{psf-compare fig} we show the results of PSF-matching in each sub-field by displaying a comparison of the enlarged PSFs to the largest PSF as a percent difference.

Due to variations in intrinsic galaxy colors and variable image depth and sky coverage, some galaxies will not appear in all bands. To get photometric measurements of all sources in every DECam image we combined the information in the five optical band images into a single detection image. We followed the procedure of \citet{szalay99} and summed the square of the signal-to-noise ratio in each band pixel-by-pixel as follows: 
\begin{equation} \label{det_map}
     D_i = \sqrt{ \sum \frac{F_{\mathrm{band}, i}^{2}}{\sigma_{\mathrm{band}, i}^{2}} },
\end{equation} 
\noindent where $D_{i}$ is the detection image $i$th pixel value, $F_{band, i}$ is the $i$th pixel flux in the $\mathrm{band}$ image, and $\sigma_{\mathrm{band}, i}$ is the rms at that $i$th pixel pulled from the the $band$ rms image. A weight image associated with this detection image was created where pixels associated with detection map pixels with data in at least one band have a value of unity and pixels associated with detection map pixels without data have a value of zero.

Photometry was measured on the PSF-matched images using the Source Extractor software (v2.19.5, \citealp{bertin96}). Catalogs were created for each of the six SHELA sub-fields with Source Extractor in two image mode using the detection image described above, and cycling through the five DECam bands as the measurement image.  In the final source catalog, we maximized the detection of faint sources while minimizing false detections by optimizing the combination of the SExtractor parameters DETECT\_THRESH and DETECT\_MINAREA. We did this by running SExtractor with an array of combinations of DETECT\_THRESH and DETECT\_MINAREA and chose the combination of 1.6 and 3, respectively, which detected all sources that appeared real by visual inspection and included the fewest false positive detections from random noise fluctuations.

We measure source colors in 1.89$\arcsec$ diameter circular apertures (which corresponds to an enclosed flux fraction of 59-75\% for unresolved sources in our sub-fields).  To obtain the total flux, we derived an aperture correction defined as the flux in a 1.89$\arcsec$ diameter aperture divided by the flux in a Kron aperture (i.e., FLUX\_AUTO), using the default Kron aperture parameters of PHOT\_AUTOPARAM= 2.5, 3.5, which has been shown to calculate the total flux to within $\sim$5\% \citep{fink15}. This correction was derived in the $r^{\prime}$-band on a per-object basis to account for different source sizes and ellipticities and was applied to the fluxes in the other DECam bands per sub-field.
In areas where the sub-fields overlapped, sources with positions that matched to within 1.2$\arcsec$ in neighboring sub-field catalogs had their fluxes mean-combined after being weighted by the inverse square of their uncertainties (see Section \ref{phot_errors}).
The DECam source fluxes were corrected for Galactic extinction using the color excess E(B-V) measurements by \citet{schlafly11}. We obtained E(B-V) values using the Galactic Dust Reddening and Extinction application on the NASA/IPAC Infrared Science Archive (IRAS) website\footnote{http://irsa.ipac.caltech.edu/applications/DUST/}. We queried IRAS for E(B-V) values (the mean value within a 5$\arcmin$ radius)  for a grid of points across the SHELA field with 4$\arcmin$ spacing and assigned each source the E(B-V) value from the closest grid point.  The \citet{cardelli89} Milky Way reddening curve parameterized by $R_V=3.1$ was used to derive the corrections at each band's central wavelength.
We compared the extinction-corrected DECam photometry to SDSS DR14 per sub-field and band and found agreement for point sources to better than $\delta m < 0.05-0.2$ mag in terms of scatter.

\input{table_err_param.tex}

\subsection{DECam Photometric Errors} \label{phot_errors}

We estimated photometric uncertainties in the DECam images by estimating the image noise in apertures as a function of pixels per aperture, N, following the procedure described in Section 2 of \citet{papovich16}. There are two limiting cases for the uncertainty in apertures with N pixels, $\sigma_N$. If pixel errors are completely uncorrelated, the aperture uncertainty scales as the square root of the number of pixels, $\sigma_N = \sigma_1 \times \sqrt{N}$, where $\sigma_1$ is the standard deviation of sky background pixels. If pixel errors are completely correlated then $\sigma_N = \sigma_1 \times N$ \citep{quadri06}. Thus, the aperture uncertainty will scale as $N^{\beta}$ with $0.5 <\beta<1$. %

\input{fig_error_fig.tex}
To estimate the aperture noise as a function of $N$ pixels, we measured the sky counts in 80,000 randomly placed apertures ranging in diameter from 0\farcs27 to 8\farcs1 across each stacked DECam image. We required apertures to fall in regions of the background sky, which we define as the region where the exposure time map has the value of the at least the median exposure time (ensuring $>$50\% of each image was considered), excluding detected sources and pixels flagged in the DQM. We also required the apertures do not overlap with each other. We then estimated $\sigma_N$ for each aperture with $N$ pixels by computing the standard deviation of the distribution of aperture fluxes from the normalized median absolute deviation, $\sigma_{\mathrm{nmad}}$ \citep{beers90}. We calculated $\sigma_1$ by computing $\sigma_{\mathrm{nmad}}$ for all pixels in the background sky as defined above. Figure \ref{error_fig} shows an example of the measured flux uncertainty in a given aperture, $\sigma_N$, as a function of the square root of the number of pixels in each aperture, $N$, for the five DECam bands in the sub-field SHELA-1.

Following \citealp{papovich16} we fit a parameterized function to the noise in an aperture of $N$ pixels, $\sigma_N$, as,
\begin{equation} \label{sigma_n}
     \sigma_N = \sigma_1 ( \alpha N^{\beta} + \gamma N^{\delta}) ,
\end{equation}  
\noindent where $\sigma_1$ is the pixel-to-pixel standard deviation in the sky background, and $\alpha$, $\beta$, $\gamma$, and $\delta$ are free parameters. The best-fitting parameters in Equation \ref{sigma_n} for the combined DECam images are listed in Table \ref{err table}. While the second term was intended to aid in fitting the data at large $N$ values, in actuality the second term contributed significantly at all $N$ values resulting in $\beta\approx$0.9-1. Nevertheless our functional fits reproduce the data well as can be seen, for example, in Figure \ref{error_fig}. To estimate how correlated the pixel-to-pixel noise is, we fit $\sigma_N$ with only the first term in Equation \ref{sigma_n} and found typical values of $\beta\approx$ 0.65-0.70 suggesting slightly correlated pixel-to-pixel noise.

The photometric errors estimated by Equation \ref{sigma_n} were scaled to apply to flux measurements outside the region with the median exposure time. The flux uncertainty for the $i$-th source in band $b$ in the sub-field $f$ is calculated as
\begin{equation} \label{final sigma}
     \sigma_{i,f,b}^{2} =  \sigma_{N,f,b}^{2} \Bigg(\frac{\mathrm{rms}_{i,f,b}}{\mathrm{rms}_{\mathrm{med},f,b}}\Bigg)^{2},
\end{equation}
\noindent where $\sigma_{N,f,b}$ is given by Equation \ref{sigma_n} for each band and sub-field, $\mathrm{rms}_{i,f,b}$ is the value of the rms map at the central pixel location of the $i$-th source in each band and sub-field, and $\mathrm{rms}_{\mathrm{med},f,b}$ is the median value of the rms map in each band and sub-field. The photometric error estimates exclude Poisson photon errors, which we estimate to contribute $<$5\% uncertainty to the optical fluxes of our high-$z$ candidates.

As described in Section \ref{sextration}, all source fluxes are measured in circular apertures of 1.89\arcsec diameter and scaled to total on a per-object basis. Likewise, the flux uncertainties in the apertures were scaled by the same amount to determine the total flux uncertainties, so that the S/N for the total flux is the same as for the aperture flux.

\subsection{IRAC data reduction and photometry} \label{irac}

\par As discussed below, we wish to enhance the validity of our $z=4$ galaxy sample by including IRAC photometry in our galaxy sample selection.  While we could allow this by position-matching the published {\it Spitzer}/IRAC catalog from \citet{papovich16} to our DECam catalog, this is not optimal for two reasons.  First, the \citet{papovich16} catalog is IRAC-detected, and so only includes sources with significant IRAC flux, while for our purposes, even a non-detection in IRAC can be useful for calculating a photometric redshift.  Second, this catalog uses apertures defined on the positions and shapes of the IRAC sources, while the larger PSF of the IRAC data results in significant blending, which is a larger issue at fainter magnitudes, where we expect to find the bulk of our sources of interest.  For these reasons, we applied the Tractor image modeling code \citep{lang16a,lang16b} to perform  ``forced photometry", which employs prior measurements of source positions and surface brightness profiles from a high-resolution band to model and fit the fluxes of the source in the remaining bands, splitting the flux in overlapping objects into their respective sources. We specifically used the Tractor to optimize the likelihood for the photometric properties of DECam sources in each of IRAC 3.6 and 4.5 $\mu$m bands given initial information on the source and image parameters.  The input image parameters of IRAC 3.6 and 4.5 $\mu$m images included a noise mode, a point spread function (PSF) model, image astrometric information (WCS), and calibration information (the ``sky noise" or rms of the image background). The input source parameters included the DECam source positions, brightness, and surface brightness profile shapes. The Tractor proceeds by rendering a model of a galaxy or a point source convolved with the image PSF model at each IRAC band and then performs a linear least-square fit for source fluxes such that the sum of source fluxes is closest to the actual image pixels, with respect to the noise model. We describe how we use the Tractor to perform forced photometry on IRAC images in detail below.

\par We begin our source modeling procedure by selecting the fiducial band high-resolution model of each source. We use the fluxes and surface brightness profile shape parameters measured in our DECam detection image because the image combines the information of all sources in the five optical band images (as described in Section \ref{sextration}).  Second, we use one-component circular Gaussian to model the PSF.  During the modeling of each source, we allow Tractor to optimize the Gaussian $\sigma$ value, in addition to optimize a source flux. We find that the median of the optimized Gaussian $\sigma$ is $0\farcs80$ (equivalent to a full-width at half maximum of $1\farcs88$) for both IRAC 3.6 and 4.5 $\mu$m images, consistent with those measured from an empirical point response function for the 3.6$\mu$m and 4.5$\mu$m image (FWHM of $1\farcs97$, see Papovich 2016, Section 3.4).

\par In practice, we extracted an IRAC image cutout of each source in the input DECam catalog. We selected the cutout size of ${16}'' \times {16}''$. This cutout size represents a trade-off between minimizing computational costs related to larger cutout sizes and ensuring that the sources lie well within the cutout extent. The sources of interest within cutout are modeled as either unresolved (i.e., a point source) or resolved based on the DECam detection image. We considered a source to be resolved if an estimated radius $r > 0\farcs1$. We define the radius as, $r_\mathrm{source} = a \times \sqrt{b/a}$, where  $a$ is a semi-major axis and $b/a$ is an axis ratio. We perform the photometry for resolved sources using a deVaucouleurs profile (equivalent to S\'{e}rsic profile with n=4) with shape parameters (semi-major axis, position angle, and axis ratio) measured using our DECam detection image. We have also performed the photometry using an exponential profile (equivalent to S\'{e}rsic profile with n=1), but we do not find any significant difference between the IRAC flux measurements for the two galaxy profiles. Therefore, we adopt a deVaucouleurs profile to model all resolved sources. The Tractor simultaneously modeled and optimized the sources of interest and neighboring sources within the cutout. Finally, the Tractor provided the measurement IRAC flux of each DECam source with the lowest reduced chi-squared value. We validated the Tractor-based IRAC fluxes by comparing the fluxes of isolated sources (no neighbors within  3$''$) to the published {\it Spitzer}/IRAC catalog from \citet{papovich16}. For both bands, we found good agreement with a bias offset of $\delta m < 0.05$ mag and a scatter of $<$0.11 mag down to $m=20.5$ mag and a bias offset of $\delta m < 0.13$ mag and a scatter of $<$0.26 down to $m=22$ mag.

%% file: fig_filters.tex
\begin{figure*}[!tbp]
\centering
\includegraphics[scale=0.6,angle=0]{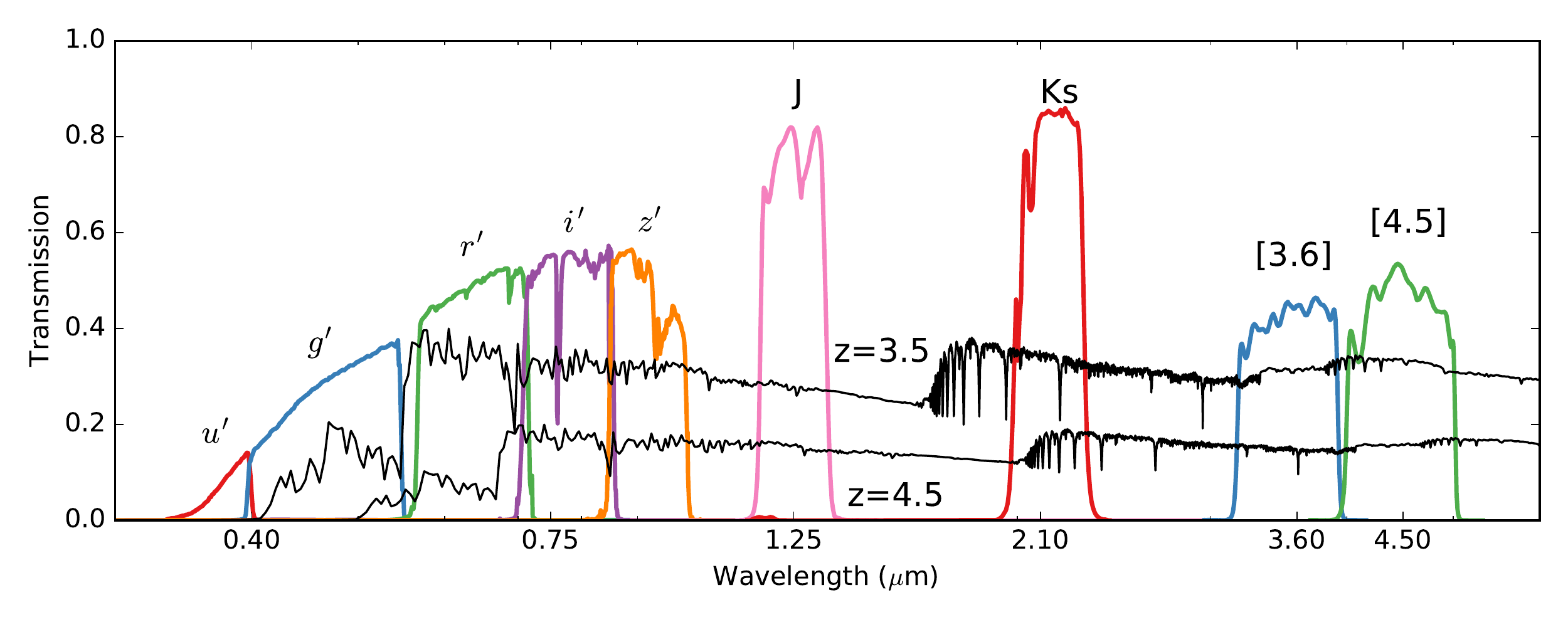}
\caption{The filter transmission curves for the nine photometric bands used in this study (curves are labelled in the figure) and two model star-forming galaxy spectra with redshifts $z=3.5$ and $z=4.5$, respectively (black). The model spectra have units of Jy and are arbitrarily scaled. The $z=3.5$ galaxy spectrum falls completely red-ward of the $u'$ band transmission curve and the $z=4.5$ galaxy spectra has almost zero flux falling in the $g'$ bandpass.}
\label{filters}
\end{figure*}

%% file: table_depth_fwhm.tex
\begin{deluxetable*}{ccc|cc|cc|cc|cc|cc}[htb!]
\tablecaption{DECam Imaging Summary - Seeing and Limiting Magnitude \label{depth_table}}
\tablecolumns{13} 
\tablehead{\colhead{} & \colhead{} & \colhead{}    &  \multicolumn{2}{c}{$u'$} &  \multicolumn{2}{c}{$g'$} &  \multicolumn{2}{c}{$r'$} &  \multicolumn{2}{c}{$i'$} &  \multicolumn{2}{c}{$z'$}   \\ 
\colhead{Sub-Field} & \colhead{R.A.} & \colhead{Dec.} & \colhead{FWHM} & \colhead{depth} & \colhead{FWHM} & \colhead{depth} & \colhead{FWHM} & \colhead{depth} & \colhead{FWHM} & \colhead{depth} & \colhead{FWHM} & \colhead{depth}  \\
\colhead{ID No.} & \colhead{(J2000)} & \colhead{(J2000)} & \colhead{($''$)} & \colhead{(mag)} & \colhead{($''$)} & \colhead{(mag)} & \colhead{($''$)} & \colhead{(mag)} & \colhead{($''$)} & \colhead{(mag)} & \colhead{($''$)} & \colhead{(mag)}
}
\startdata
SHELA-1 & $1^{\textrm{h}}00^{\textrm{m}}52.8^{\textrm{s}}$ & -0\arcdeg00\arcmin36\arcsec & 1.12 & 25.2 & 1.06 & 24.8 & 1.0 & 24.8 & 0.87 & 24.5 & 0.86 & 23.8 \\
SHELA-2 & $1^{\textrm{h}}07^{\textrm{m}}02.4^{\textrm{s}}$ & -0\arcdeg00\arcmin36\arcsec & 1.2 & 25.2 & 1.26 & 24.8 & 1.28 & 24.6 & 1.39 & 23.9 & 0.96 & 23.6 \\
SHELA-3 & $1^{\textrm{h}}13^{\textrm{m}}12.0^{\textrm{s}}$ & -0\arcdeg00\arcmin36\arcsec & 1.22 & 25.4 & 1.36 & 25.0 & 1.14 & 24.7 & 1.04 & 24.5 & 1.21 & 23.5 \\
SHELA-4 & $1^{\textrm{h}}19^{\textrm{m}}21.6^{\textrm{s}}$	& -0\arcdeg00\arcmin36\arcsec &  1.15 & 25.3 & 1.4 & 24.4 & 1.05 & 24.3 & 1.0 & 22.1 & 1.13 & 23.7 \\
SHELA-5 & $1^{\textrm{h}}25^{\textrm{m}}31.2^{\textrm{s}}$ & -0\arcdeg00\arcmin36\arcsec & 1.21 & 25.1 & 1.07 & 24.9 & 1.02 & 24.3 & 0.93 & 23.9 & 0.85 & 23.6 \\
SHELA-6 & $1^{\textrm{h}}31^{\textrm{m}}40.8^{\textrm{s}}$ & -0\arcdeg00\arcmin36\arcsec & 1.26 & 25.4 & 1.37 & 24.9 & 1.26 & 24.5 & 1.27 & 24.2 & 0.86 & 23.5 \
\enddata
\tablecomments{The FWHM values are for the stacked DECam images before PSF matching and have units of arcseconds. The magnitudes quoted are the 5-$\sigma$ limits measured in 1.89"-diameter apertures on the PSF-matched images (see Section \ref{sextration}).}
\end{deluxetable*}

%% file: fig_psf-compare.tex
\begin{figure*}[!tbp]
\includegraphics[scale=0.6,angle=0]{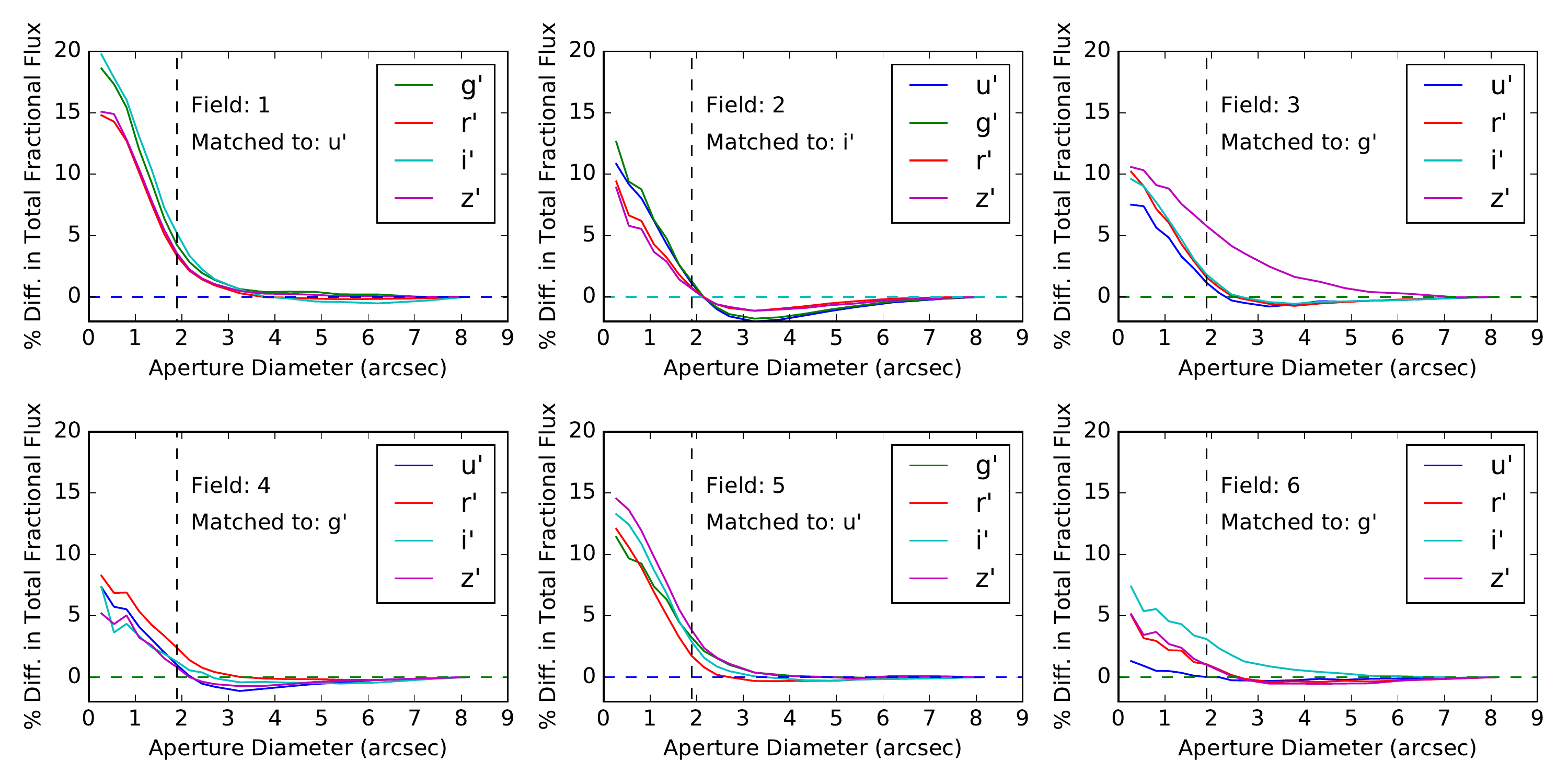}
\caption{The results of PSF-matching the DECam images. Each panel shows the percent difference between the enlarged PSFs and the largest PSF per SHELA sub-field. The colored lines correspond to the four bands listed in each panel's legend. The vertical dashed line denotes 1.89" which is the aperture diameter at which we compared PSFs during the PSF-matching procedure (see Section \ref{sextration} for details). The horizontal dashed line was placed to zero to guide the eye. This figure illustrates that all bands in all sub-fields have PSFs that collect the same fraction of light as their respective largest PSF to within 5\% except for the $z'$ band in sub-field 3 which matches to about 6\%.}
\label{psf-compare fig}
\end{figure*}

%% file: table_err_param.tex
\begin{deluxetable}{cccccccc}[ht]
\tablecaption{Fit Parameters for Background Fluctuations as Function of Aperture Size Using Eq.(\ref{sigma_n})\label{err table}}
\tablehead{\colhead{Sub-Field ID} & \colhead{Band} & \colhead{$\sigma_{1}$} & \colhead{$\alpha$} & \colhead{$\beta^{\dagger}$}  & \colhead{$\gamma$} & \colhead{$\delta$} & \colhead{$\mathrm{rms}_{\mathrm{med}}$} \\
\colhead{} & \colhead{} & \colhead{(nJy)} & \colhead{} & \colhead{}  & \colhead{} & \colhead{} & \colhead{(nJy)}
}
\startdata
SHELA-1 & $u^{\prime}$ & 5.02 & 0.09 & 0.89 & 1.33 & 0.35 & 5.58 \\
 & $g^{\prime}$ & 4.07 & 0.21 & 0.83 & 1.71 & 0.43 & 8.85 \\
 & $r^{\prime}$ & 4.14 & 0.18 & 0.91 & 2.37 & 0.36 & 10.2 \\
 & $i^{\prime}$ & 4.1 & 0.13 & 1.0 & 2.88 & 0.42 & 14.5 \\
 & $z^{\prime}$ & 7.16 & 0.19 & 0.93 & 2.55 & 0.45 & 25.0 \\
\hline 
SHELA-2 & $u^{\prime}$ & 1.21 & 0.24 & 0.91 & 2.41 & 0.55 & 5.88 \\
 & $g^{\prime}$ & 1.8 & 0.35 & 0.87 & 2.3 & 0.5 & 8.26 \\
 & $r^{\prime}$ & 3.02 & 0.21 & 0.94 & 2.68 & 0.41 & 10.2 \\
 & $i^{\prime}$ & 14.3 & 0.05 & 1.0 & 1.37 & 0.36 & 15.8 \\
 & $z^{\prime}$ & 5.36 & 0.34 & 0.9 & 2.48 & 0.51 & 27.7 \\
\hline 
SHELA-3 & $u^{\prime}$ & 1.68 & 0.23 & 0.89 & 2.2 & 0.42 & 4.81 \\
 & $g^{\prime}$ & 5.36 & 0.12 & 0.88 & 1.33 & 0.33 & 6.25 \\
 & $r^{\prime}$ & 2.77 & 0.19 & 0.94 & 2.63 & 0.45 & 9.8 \\
 & $i^{\prime}$ & 3.06 & 0.32 & 0.92 & 3.07 & 0.38 & 12.6 \\
 & $z^{\prime}$ & 11.0 & 0.15 & 0.91 & 1.95 & 0.41 & 24.1 \\
\hline 
SHELA-4 & $u^{\prime}$ & 1.08 & 0.26 & 0.95 & 2.54 & 0.49 & 4.64 \\
 & $g^{\prime}$ & 7.36 & 0.13 & 0.85 & 1.24 & 0.34 & 8.31 \\
 & $r^{\prime}$ & 2.86 & 0.25 & 0.91 & 2.44 & 0.51 & 12.1 \\
 & $i^{\prime}$ & 24.9 & 1.42 & 0.6 & 1.27 & 0.6 & 63.5 \\
 & $z^{\prime}$ & 5.48 & 0.26 & 0.93 & 2.79 & 0.45 & 22.6 \\
\hline 
SHELA-5 & $u^{\prime}$ & 5.25 & 0.11 & 0.87 & 1.25 & 0.33 & 5.8 \\
 & $g^{\prime}$ & 3.11 & 0.23 & 0.87 & 1.92 & 0.43 & 7.86 \\
 & $r^{\prime}$ & 4.78 & 0.38 & 0.81 & 2.02 & 0.46 & 15.8 \\
 & $i^{\prime}$ & 5.53 & 0.33 & 0.85 & 2.32 & 0.46 & 21.9 \\
 & $z^{\prime}$ & 6.86 & 0.23 & 0.92 & 2.53 & 0.49 & 28.0 \\
\hline 
SHELA-6 & $u^{\prime}$ & 1.66 & 0.15 & 0.95 & 2.38 & 0.42 & 4.6 \\
 & $g^{\prime}$ & 6.15 & 0.12 & 0.86 & 1.27 & 0.32 & 6.8 \\
 & $r^{\prime}$ & 4.81 & 0.21 & 0.87 & 2.09 & 0.4 & 12.1 \\
 & $i^{\prime}$ & 6.87 & 0.14 & 0.93 & 2.14 & 0.37 & 14.7 \\
 & $z^{\prime}$ & 6.09 & 0.51 & 0.82 & 2.13 & 0.53 & 32.8 \
\enddata  
\tablecomments{$^\dagger$Typical values of $\beta\approx$ 0.65-0.70 when using a two parameter fit (i.e. with only $\alpha$ and $\beta$) suggest slightly correlated noise between pixels.}
\end{deluxetable}

%% file: fig_error_fig.tex
\begin{figure}[h]
\centering
\includegraphics[scale=0.7,angle=0]{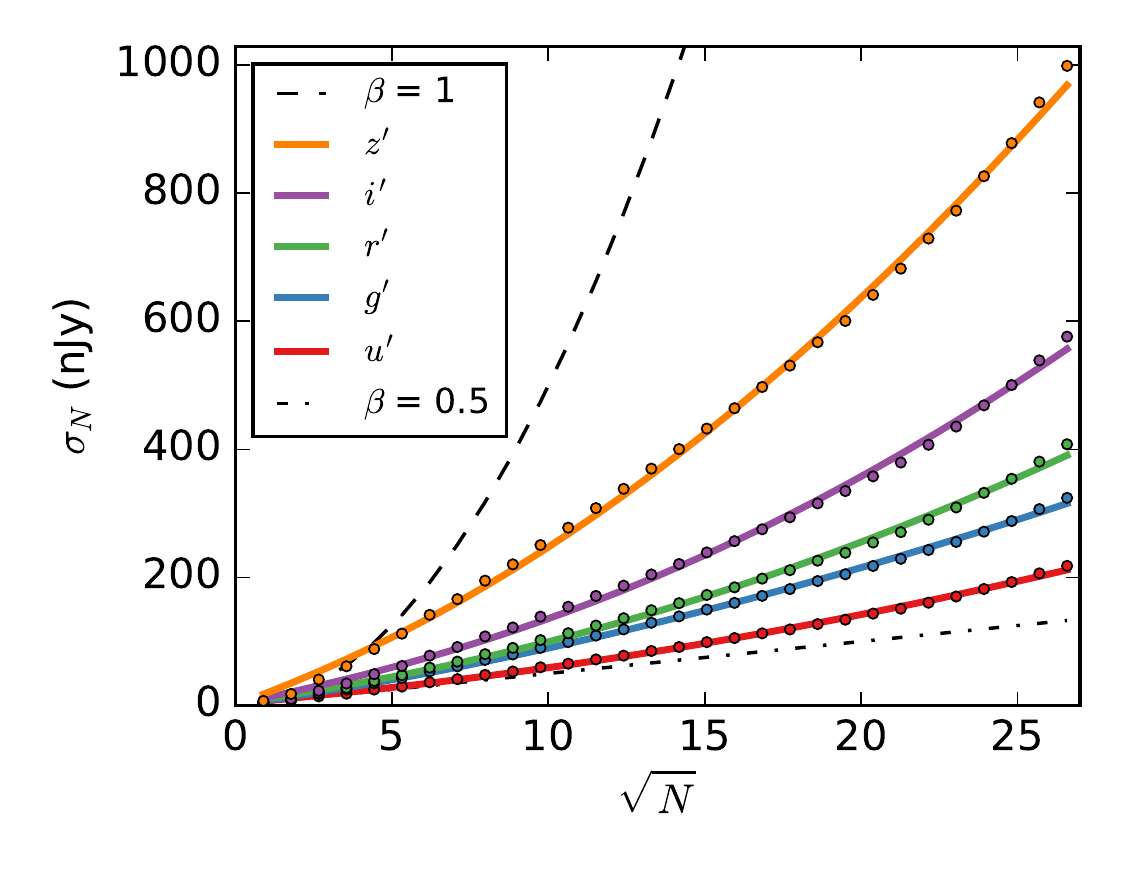}
\caption{Background noise fluctuations, $\sigma_{N}$, in an aperture of $N$ pixels plotted as a function of the square root of the number of pixels for the five DECam band images in sub-field SHELA-1. The colored dotted lines are the measured aperture fluxes and the solid lines are fits to the data. See legend insert for color coding information. The dot-dashed line shows the relation assuming uncorrelated pixels, $\sigma_{N} \sim \sqrt{N}$ . The dashed line shows the relation assuming perfectly correlated pixels ($\sigma_{N} \sim N$; \citealp{quadri06}).}
\label{error_fig}
\end{figure}

%% file: s3_sample.tex
\section{Sample Selection} \label{sample}
\subsection{Photometric Redshifts and Selection Criteria} \label{selection}
We selected our sample of high-redshift galaxies using a selection procedure that relies on photometric redshift ($z_{\mathrm{phot}}$) fitting, which leverages the combined information in all photometric bands used. We obtained $z_{\mathrm{phot}}$'s and $z_{\mathrm{phot}}$ probability distribution functions (PDFs) from the EAZY software package \citep{brammer08}. For this analysis, we use the ``z\_a'' redshift column from EAZY which is produced by minimizing the $\chi^2$ in the all-template linear combination mode. We also tried the ``z\_peak'' column and our resulting luminosity function did not change significantly. EAZY assumes the intergalactic medium (IGM) prescription of \citet{madau95}. We did not use any magnitude priors based on galaxy luminosity functions when running EAZY as the existing uncertainties in the bright-end would bias our results (effectively assuming a flat prior). We then applied a number of selection criteria using the $z_{\mathrm{phot}}$ PDFs from EAZY following the procedure of \citet{fink15}, which we summarize here, to construct a $z \approx$ 4 galaxy sample. First, we required a source to have a signal-to-noise ratio of greater than 3.5 in the two photometric bands ($r'$ and $i'$ bands) that probe the UV continuum of our galaxy sample, which has been shown to limit contamination by noise to negligible amounts by \citet{fink15}. Next we required the integral of each source's $z_{\mathrm{phot}}$ PDF for $z>2.5$ to be greater than 0.8 and the integral of the $z_{\mathrm{phot}}$ PDF from $z=$ 3.5-4.5 to be greater than the integral of the $z_{\mathrm{phot}}$ PDF in all other $\Delta z=1$ bins centered on integer-valued redshifts. Finally, we required a source to have no photometric flags on its $u'$, $g'$, $r'$, and $i'$ flux measurements in our photometric catalog. These four bands are crucial for probing both sides of the Lyman break of galaxies in the redshift range of interest. Photometric flags indicate saturated pixels, transient sources, or bad pixels as defined by the CP (see section \ref{data_reduction}).

Initially, we required sources to pass this $z_{\mathrm{phot}}$ selection procedure when only the DECam and IRAC bands were used. After inspecting candidates from this selection and finding low-$z$ contaminants (see Section \ref{contamination}) we moved to include the VISTA $J$ and $K$ photometry when available. The candidate sample of high-$z$ galaxies after including the VISTA $J$ and $K$ photometry significantly reduced the amount of obvious low-$z$ contamination as described in the following section, however, a new class of contaminant appeared in the candidate sample due to erroneous VISTA photometry for some sources. To overcome this we required a source to pass the $z_{\mathrm{phot}}$ selection process with and without including the VISTA $J$ and $K$ photometry. This double $z_{\mathrm{phot}}$ selection process resulted in an initial sample of 4,364 high-$z$ objects. Next, we performed an investigation into possible contamination, which resulted in additional selection criteria and a refined sample.

\subsection{Identifying Contamination} \label{contamination}

Photometric studies of high-$z$ objects can be contaminated by galactic stars and low-$z$ galaxies whose 4000 \AA\ break can mimic the Lyman-$\alpha$ break of high-$z$ galaxies. The inclusion of IRAC photometry is crucially important for removing galactic stars from our sample as galactic stars have optical colors very similar to $z=4$ objects (Figure \ref{fig:cc}). While inspecting the photometry and best-fitting SED templates to confirm our fits were robust and our high-$z$ galaxy candidates were convincing we found evidence of contamination in our preliminary photometrically selected sample. We explored ways of identifying and removing the contamination including cross-matching our sample with proper motion catalogs, SDSS spectroscopy, and X-ray catalogs. Additionally, we implemented machine learning methods.

\subsubsection{Cross-matching with NOMAD and SDSS} \label{contam:nomad and sdss}
While inspecting the photometry of our preliminary sample derived before the inclusion of the VISTA data we found a fraction of candidates had red $r'-i'$ colors and excesses in the $i'$ and $z'$ bands with respect to the best fitting $z\sim$ 4 template. We investigated whether these objects had low redshift origins by cross-matching our sample with the Naval Observatory Merged Astrometric Dataset (NOMAD) proper motion catalog \citep{Zacharias04} and the SDSS spectral catalog (DR13; \citealt{abareti17}). The cross-matching with NOMAD resulted in the identification of 16 objects with proper motion measurements, 6 of which were large in magnitude suggesting that these sources were stellar contaminants. Cross-matching with SDSS spectra resulted in the identification of 23 $z\sim$ 4 QSOs, two low mass stars, and one low-$z$ galaxy in our sample. All of the stellar objects and the low-$z$ galaxy had red $r'-i'$ colors confirming our suspicion that a fraction of our sample had low-$z$ origins. We removed from our candidate sample the 6 objects with proper motions above 50 mas/yr in NOMAD. We chose the threshold 50 mas/yr because some SDSS QSOs are reported to have small nonzero proper motions in the NOMAD catalog. After inclusion of the VISTA $J$ and $K$ photometry, 5 of the 6 rejected objects became best-fit by a low-$z$ solution and therefore were rejected by our selection process automatically. Results like this suggested the inclusion of the VISTA data improved the fidelity of our sample.

\subsubsection{Cross-matching with X-ray catalog Stripe 82X} \label{contam:xray}
In principle, AGNs can be distinguished from UV-bright star-forming galaxies at high $z$ by their X-ray emission as AGNs dominate X-ray number counts down to $F_X = 1 \times 10^{-17} \mathrm{erg\ cm}^{-2}\ \mathrm{s}^{-1}$ in 0.5-2 keV \textit{Chandra} imaging \citep{lehmer12}. In fact, \citet{giallongo15} found 22 AGN candidates at $z>4$ by measuring fluxes in deep 0.5-2 keV \textit{Chandra} imaging at the positions of their optically-selected candidates. They required AGN candidates to have an X-ray detection of $F_X > 1.5 \times 10^{-17}~\mathrm{erg\ cm}^{-2}\ \mathrm{s}^{-1}$. We attempted to identify AGNs in our candidate sample by position matching with the 31 deg$^2$ Stripe 82X X-ray catalog \citep{lamassa16}. Unfortunately, the flux limit at 0.5-2 keV was $8.7 \times 10^{-16}~\mathrm{erg\ cm}^{-2}\ \mathrm{s}^{-1}$, shallower than the detection threshold used by \citet{giallongo15}. Cross-matching the Stripe 82X X-ray Catalog with our candidate sample resulted in 8 matches within 7$''$ or the matching radius used by \citet{lamassa16} to match the Stripe 82X X-ray Catalog with ancillary data. Two of the X-ray-matched sources ($m_{i'}\approx$19) are SDSS AGNs, although one has two extended objects within the 7$''$ matching radius, drawing into question the likelihood of the match. Another X-ray-matched source ($m_{i'}\approx$22) is a very red object ($r'-i'=-$0.9 and $i'-[3.6]=$3.4) without SDSS spectroscopy. The remaining five X-ray-matched sources are fainter ($m_{i'}\approx$24) and without SDSS spectroscopy, and two have large separations ($>6''$) with their X-ray counterpart. All 8 X-ray-matched sources satisfied each of our selection criteria and made it into our final sample.

\subsubsection{Insights from machine learning} \label{contam:ml}
To further understand and minimize the contamination in our preliminary sample we utilized two machine learning algorithms: a decision tree algorithm and a random forest algorithm. First, we tried the decision tree algorithm from the sci-kit learn Python package \citep{scikit-learn}. Executing the decision tree algorithm involved creating a training set by classifying (via visual inspection) the 300 brightest objects as one of five types: 1. obviously high-$z$ galaxy or high-$z$ AGN, 2. obviously low-$z$ galaxy, 3. indistinguishable between high-$z$ or low-$z$ object, 4. SDSS spectroscopically identified QSO, 5. spurious source. The classification was driven by a combination of the shape of the optical SED, the $\chi^2$ of the best fitting high-$z$ ($z\sim$ 4) galaxy template, the difference in $\chi^2$ between the best fitting high-$z$ galaxy template and the best fitting low-$z$ galaxy template, SDSS spectral classification, and proper motion measurements. The obviously high-$z$ objects and the SDSS QSOs had roughly flat rest-UV spectral slopes (i.e., the $i'$, $r'$, and $z'$ fluxes were comparable) while the obviously low-$z$ objects had a clear spectral peak between the optical and mid-IR, specifically the $r'-i'$ and $i'-z'$ colors were quite red while the $z'-[3.6]$ was blue. We then selected three data "features" or measurements for the decision tree to choose from to predict the classifications of the training set: 1. The $\chi^2$ of the best fitting high-$z$ galaxy template, 2. The $r'-i'$ color, and 3. The signal-to-noise ratio in the $u'$ band. We include this signal-to-noise ratio because most obviously high-$z$ sources had a signal-to-noise ratio of less than 3 (as they should, as this filter is completely blue-ward of the Lyman break for the target redshift range) while the obviously low-$z$ source did not. We restricted the training set to include only the obviously high-$z$ objects (including SDSS classified QSOs) and obviously low-$z$ objects.

To evaluate the decision tree performance we assigned 34\% of the training set to a test set and left the test set out of the first round of training. The first round of training was validated using three-fold cross-validation with a score of 94+/-2\%. Three-fold cross validation verifies the model is not over-fitting or overly dependent on a small randomly selected training (or validation) set. The process of three-fold cross-validation involves splitting the training sample into three sub-samples, training the model independently on each combination of two sub-samples while validating on the remaining sub-sample, and then averaging the validation scores of all the combinations. A validation score of 100\% indicates each sub-sample combination trained a model that successfully predicted the classifications of every object in the remaining sub-sample. After validation, we applied the model to the test set and achieved a test score of 91\%. We then incorporated the test set into the training set and retrained the model. This second round of training had a 3-fold cross validation score of 94 $\pm$ 5\%. The algorithm determined that classification of the training set could be predicted at 92\% accuracy using the $r'-i'$ color and the $\chi^2$ only, where obviously high-$z$ objects have $r'-i'$ $<$ 0.555 and $\chi^2$ $<$ 70.6.

After performing the decision tree machine learning we tried the random forest algorithm by using the scikit-learn routine RandomForestClassifier \citep{scikit-learn}. The random forest algorithm uses the combined information of all the features of our dataset instead of using only the three well-motivated features that we provided the decision tree algorithm. The random forest algorithm fits a number of decision tree classifiers on a randomly drawn and bootstrapped subset of the data using a randomly drawn subset of the dataset features. The decision tree classifiers are averaged to maximize accuracy and control over-fitting. To provide the best classifications to the algorithm we re-inspected the photometry and the besting-fitting EAZY template SEDs of the brightest 311 objects ($m_{\mathrm{i}'} < 22.9$) and classified by eye each with a probability of being at high $z$, at low $z$, and a spurious source. We also inspected the best-fitting EAZY template at the redshift of the second highest peak in the redshift PDF, which was usually between $z=0.1-0.5$. The features we provided the algorithm included all colors using the DECam and IRAC photometric bands, the $\chi^2$ value of the best-fitting EAZY template, and the u-band S/N ratio. We ran the random forest algorithm using 1000 estimators (or decision trees) with a max depth of two and balanced class weights. While the classification accuracy of the random forest was only marginally better than the decision tree, we were able to determine the relative importance of the features within the random forest. The most important feature was the $i'-[3.6]$ color, which was consistent with our prior observations of the low-$z$ contaminants having blue $z'-[3.6]$ while the SDSS classified AGNs had flat or red $z'-[3.6]$. We compared the predictive power of the random forest algorithm to that of a simple $i'-[3.6]$ color cut--where high-$z$ candidates were required to have an $i'-[3.6]>-0.2$--and found the simple color cut to be just as predictive. We, therefore, elected to adopt the simple $i'-[3.6]$ color cut as an additional step in our selection process. After we applied this cut our high-$z$ object sample consisted of 3,833 objects. We then inspected the brightest 3,200 brightest candidates ($m_{\mathrm{i}'}<24.0$) and removed 61 spurious sources cutting our candidate sample to 3,772 objects. The final step in our candidate selection process was an $r'-i'$ color cut, which was a result of the contamination test described in the following subsection.

\subsection{Estimating Contamination Using Dimmed Real Sources} \label{est contam}
We estimated the contamination in our high-$z$ galaxy sample by simulating faint and low-S/N low-$z$ interloper galaxies following a procedure used by \citet{fink15}. We did this by selecting a sample of bright low-$z$ sources from our catalog, dimming and perturbing their fluxes, and assigned the appropriate uncertainties to their dimmed fluxes. This empirical test assumes that bright low-$z$ galaxies have the same colors as the faint lower-$z$ galaxies. The sources we dimmed all had $m_{r'} =$ 14.3 - 18 mag, brighter than our brightest $z =$3.5-4.5 candidate source, a best-fitting $z_{\mathrm{phot}}$ of 0.1 $<z_{\mathrm{phot}}<$ 0.6, and no photometric flags in any optical band (see section \ref{data_reduction} for definition of photometric flag). We reduced the source fluxes randomly to create a flat distribution of dimmed $r'$-band mag spanning the range of our candidate high-$z$ galaxies ($m_{r'} =$ 18-26). We assigned flux uncertainties to the dimmed fluxes by randomly drawing from the flux uncertainties of our candidate high-$z$ galaxy sample. We then perturbed the dimmed fluxes simulating photometric scatter by drawing flux perturbations from a Gaussian distribution with a standard deviation equal to the assigned flux uncertainties. Since the VISTA $J$ and $K$ photometry from the VICS82 Survey covered $\sim$85\% of the total SHELA survey area, $\sim$15\% of our high-$z$ galaxy candidates do not have flux measurements in the VISTA bands. We incorporated this property of our catalog into the mock catalog by randomly omitting dimmed $J$ and $K$ fluxes at the same rate as the fraction of missing $J$ and $K$ fluxes in our final sample for a given $m_{r'}$ bin.

We created 4.8 million artificially dimmed sources and ran them through EAZY and our selection criteria in an identical manner as our real catalog. We found 30,594 artificially dimmed sources satisfied our $z_{\mathrm{phot}} = 4$ selection criteria. We inspected the undimmed and unperturbed properties of these sources and found a range of $r'-i'$ colors (0 $< r'-i' < 1.4$), and the objects with the reddest $r'-i'$ colors contaminated at the highest rate. We defined the contamination rate in the $i$-th $r'-i'$ color bin as

\begin{equation}
R_{i} = \frac{N_{\textrm{dimmed,selected,}i}}{N_{\textrm{dimmed,}i}},
\label{contam rate}
\end{equation}
\noindent where $N_{\textrm{dimmed,selected,}i}$ is the number of dimmed sources satisfying our $z_{\mathrm{phot}}=4$ criteria in the $i$-th $r'-i'$ color bin and $N_{\textrm{dimmed,}i}$ is the total number of dimmed sources in the $i$-th $r'-i'$ color bin. We created seven $r'-i'$ color bins spanning the range of $r'-i'$ colors recovered (0 $< r'-i' <$ 1.4) with each bin having width=0.2 mag. The contamination fraction, $F$, in our high-$z$ galaxy sample was defined as 

\begin{equation}
F = \frac{\sum_{i}^{ }{ \frac{R_i \times N_{\textrm{total},i}}{(1-R_i)}}}{N_z},
\label{contam frac}
\end{equation}
\noindent where $R_i$ is the contamination rate defined by Equation \ref{contam rate}, $N_{\textrm{total},i}$ is the total number of sources in our object catalog in the i-th $r'-i'$ color bin with $0.1 < z_{\mathrm{phot}} < 0.6$ and $N_{z}$ is the number of high-$z$ candidates in our final sample in a given redshift bin. We calculated $F$ as a function of $m_{r'}$ and $m_{i'}$ and found contamination fractions of between 0-20\% generally increasing as a function of magnitude and not exceeding 25\% in any magnitude bin above our 50\%-completeness limit ($m_{i'} > 23.5$). We learned we can improve our fidelity by implementing a $r'-i'<1.0$ color cut, given that the objects with the reddest $r'-i'$ colors contaminated at the highest rate. We then re-ran our contamination simulation with our final set of selection criteria and found improvement by a few percentage points in two of our brighter magnitude bins ($m_{i'}=$ 18.75, 20.5) and, again, contamination fractions of between 0-20\% generally increasing as a function of magnitude and not exceeding 25\% in any magnitude bin brighter than our 50\%-completeness limit as shown in Figure \ref{contam fig}. The addition of the $r'-i'<1.0$ color cut reduced our high-$z$ candidate sample size from 3,772 to the final size of 3,740 with a median $z_{\mathrm{phot}}$ of 3.8. The measured $i'$-band magnitude and the redshift distribution of our final sample is shown in Figure \ref{mag_z_histo}. A summary of our sample selection criteria is listed in Table \ref{tab:selection}. Bright ($m_{i'}<22$) candidates are plotted in Figure \ref{fig:cc} showing the distinct color parameter space they occupy (along with SDSS spectroscopically classified AGNs) compared to the parameter space occupied by  SDSS spectroscopically classified stars in the SHELA field.

\input{table_selection_rules.tex}

\input{fig_mag_z_histo.tex}

\input{fig_cc_both.tex}

\input{fig_contam_fig.tex}

\subsection{Completeness Simulations} \label{completeness section}
A crucial component of calculating the UV luminosity function is measuring the effective volume of the survey. The effective volume measurement depends on quantifying the survey incompleteness due to image depth and selection effects. To quantify the survey incompleteness we simulated a diverse population of high-$z$ galaxies with assigned photometric properties and uncertainties consistent with our source catalog and measured the fraction of simulated sources that satisfied our high-$z$ galaxy selection criteria as a function of the absolute magnitude and redshift.

The simulated mock galaxies were given properties drawn from distributions in redshift and dust attenuation (e.g., E[B-V]) while the ages and metallicities were fixed at 0.2 Gyr and solar ($Z=\textup{Z}_{\odot}$), respectively. Because the fraction of recovered galaxies per redshift bin is independent of the number of simulated galaxies per redshift bin as long as low-number statistics are avoided, the redshift distribution was defined to be flat from $2 < z < 6$, and the E(B-V) distribution was defined to be log-normal spanning $0 <$ E(B-V) $< 1$ and peaking at 0.2. Mock SEDs were then generated for each galaxy using pythonFSPS\footnote{http://dfm.io/python-fsps/current/} (Foreman-Mackey et al.\, 2014), a python package that calls the Flexible Stellar Population Synthesis Fortran library \citep{Conroy09, Conroy10}. We then integrated each galaxy SED through our nine filters (DECam $u^{\prime}$, $g^{\prime}$, $r^{\prime}$, $i^{\prime}$, and $z^{\prime}$; VISTA $J$ and $K$; and IRAC 3.6 $\mu$m and 4.5 $\mu$m). Each set of mock photometry was then scaled to have an r-band apparent magnitude within a log-normal distribution spanning 18 $< m_{r'} <$ 27. This distribution ensured we were simulating the most galaxies at the fainter magnitudes where we expected to be incomplete and fewer at bright magnitudes where we expected to be very complete. Flux uncertainties and missing VISTA fluxes were assigned in the same way as during the contamination test in Section \ref{est contam}. Mock galaxies were not assigned photometric flags. This likely results in an incompleteness of only a couple percent as the fraction of all sources affected by flags is small. All sources with a signal-to-noise ratio of greater than 3.5 in any one band, are flagged in another band less than 2\% of the time on average (never exceeding 4\%). In addition all images have $\sim$0.5\% of all pixels flagged on average (never exceeding 2\%).

The photometric catalog of 600,000 mock galaxies was then run through EAZY to generate $z_{\mathrm{phot}}$ PDFs and our high-$z$ galaxy selection was applied. The completeness was defined as the number of mock galaxies recovered divided by the number of input mock galaxies, as a function of input absolute magnitude and redshift. Figure \ref{completeness fig} shows the results of our simulation.  We define the 50\%-completeness limit as the absolute magnitude where the area under the curve falls to less than 50\% the areas under the average of the $M_{UV,i'}=-25$ to $M_{UV,i'}=-28$ curves, which we find to be at  $M_{UV,i'}=-22$ ($m =$ 24 for $z =$ 4).

\input{fig_completeness.tex}

%% file: table_selection_rules.tex
\begin{deluxetable}{cc}\centering
\tablecaption{Summary of $z=4$ Sample Selection Criteria\label{tab:selection}}
\tablehead{\colhead{Criterion} & \colhead{Section Reference}}
\startdata
S/N$_{r'} > 3.5$ & (\ref{selection}) \\
S/N$_{i'} > 3.5$ & (\ref{selection}) \\
$\int_{2.5}^{\infty} \mathrm{PDF}(z) \ dz> 0.8$ & (\ref{selection}) \\
$\int_{3.5}^{4.5} \mathrm{PDF}(z) \ dz>$\ All other $\Delta z=1$ bins & (\ref{selection}) \\
$i'-[3.6] > -0.2$ & (\ref{contam:ml}) \\
$r'-i' < 1.0$  & (\ref{est contam}) \
\enddata  
\end{deluxetable}

%% file: fig_mag_z_histo.tex
\begin{figure}[h]
\includegraphics[scale=0.7,angle=0]{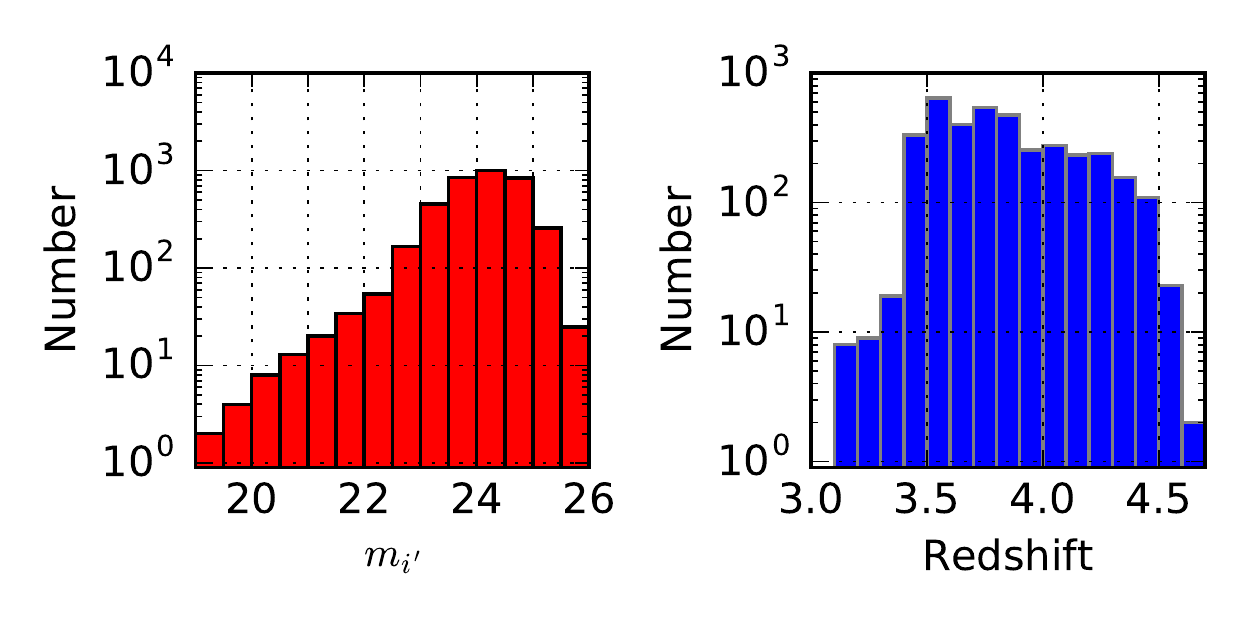}
\caption{The $i'$-band magnitude distribution (left) and the photometric redshift distribution (right) of our final sample of $z \approx$ 4 candidates. Photometric redshifts are the redshifts where $\chi^2$ is minimized for the all-template linear combination mode from the EAZY software (z\_a). We have found high-$z$ sources over a wide range of brightnesses and across the entire $z=$3.5-4.5 range.}
\label{mag_z_histo}
\end{figure}

%% file: fig_cc_both.tex
\begin{figure*}[!tbp]
\includegraphics[scale=0.72,angle=0]{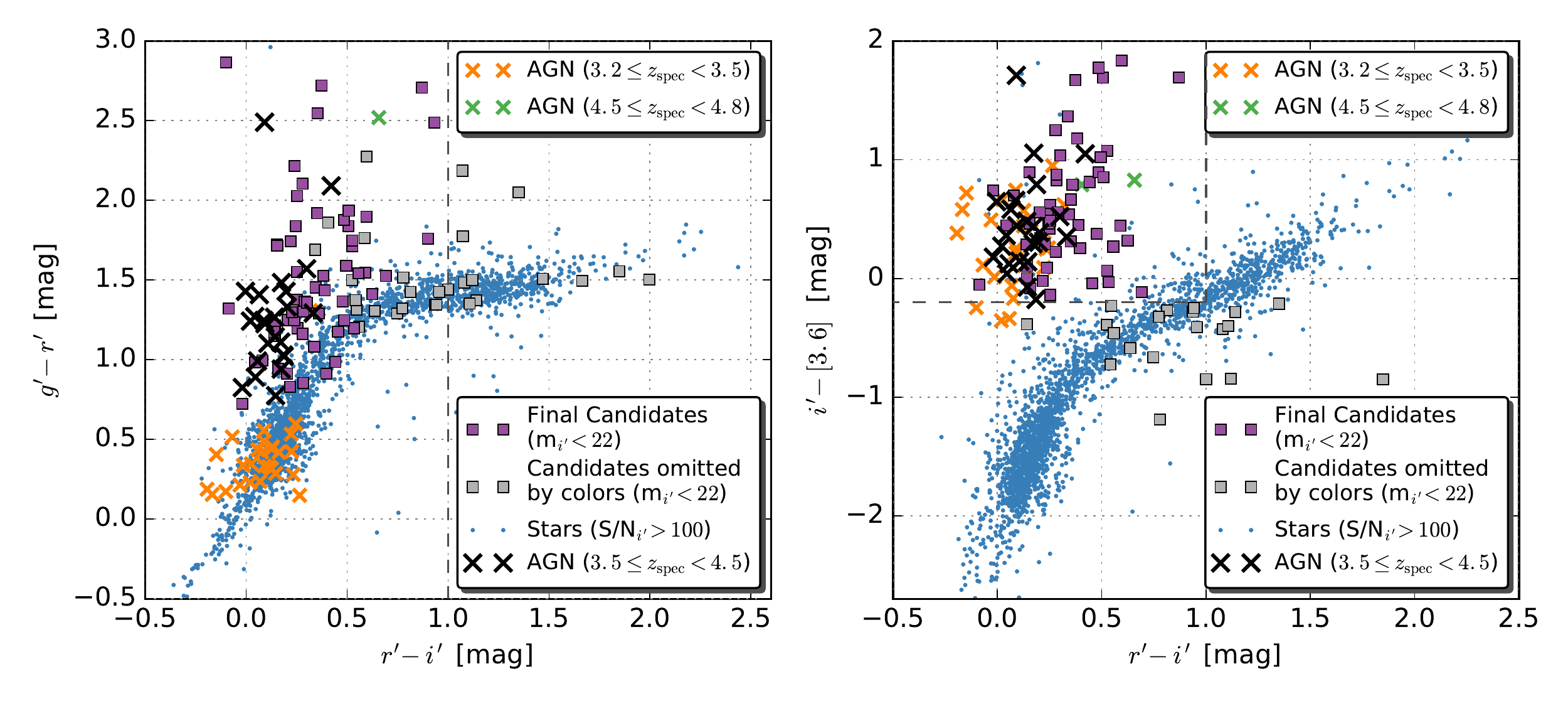}
\caption{Left: Color-color plot showing DECam $g'-r'$ vs $r'-i'$. Blue points correspond to bright ($S/N_{i} > 100$) sources classified as stars within SDSS DR14. Sources spectroscopic identified as QSO within SDSS DR13 at $3.2\leq z<3.5$, $3.5\leq z<4.5$, and $4.5\leq z < 4.8$ are orange circles, black ``x"s, and green circles, respectively. Bright candidate objects from our study are shown as squares (see legend insert for color coding). Right: Color-color plot showing $i'-[3.6]$ vs $r'-i'$. The $r'-i<1.0$ and the $i'-[3.6]>-0.2$ color selection criteria are denoted as the vertical dashed line and the horizontal dashed line, respectively. No candidates are hidden by the legend inserts. These plots illustrate how the inclusion of IRAC photometry breaks the optical color degeneracy of $z=4$ sources and galactic stars. In the left panel where only optical colors are used the $z\sim4$ AGNs share the parameter space of galactic stars, while in the right panel where one color includes the $[3.6]$ IRAC band there is a larger separation.}
\label{fig:cc}
\end{figure*}

%% file: fig_contam_fig.tex
\begin{figure}[!tbp]
\includegraphics[scale=0.575,angle=0]{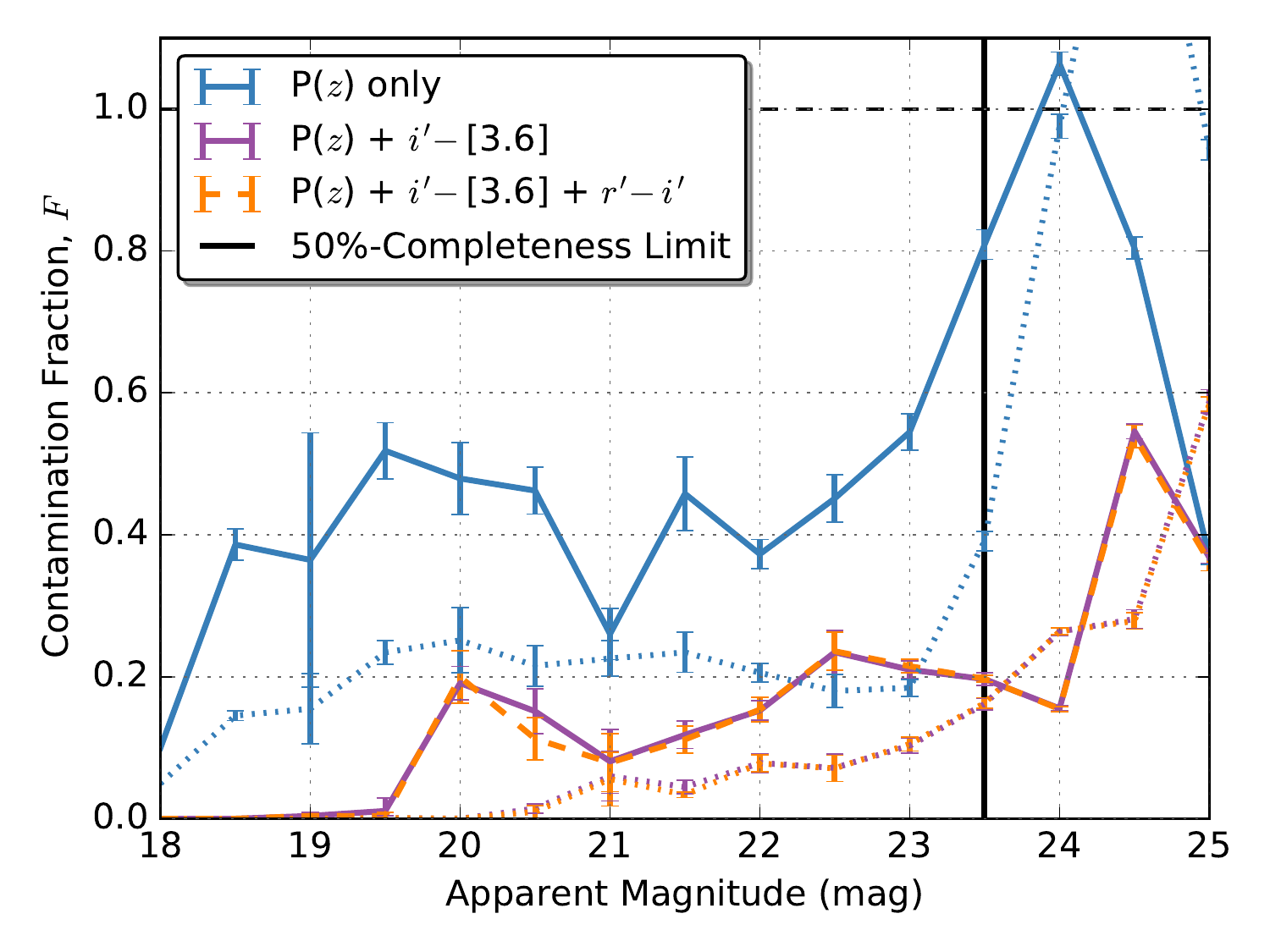}
\caption{The results of our contamination simulations estimating the contamination fraction using dimmed real sources, showing the contamination fraction, $F$, as a function of $m_{i'}$ (solid and dashed colored lines) and $m_{r'}$ (dotted colored lines). The color of the lines represent the selection criteria applied before $F$ was calculated: blue for $F$ after only the $z_{phot}$ PDF selection cuts, purple for $F$ after the $z_{phot}$ PDF selection and the $i'-[3.6]$ color cuts, and orange for $F$ after the $z_{phot}$ PDF selection, the $i'-[3.6]$ color, and the $r'-i'$ color cuts. The error bars indicate the standard deviation of the mean from subdividing our 4.8 million simulated sources into 20 subsamples. The final contamination fraction was found to be between 0-20\% generally increasing as a function of magnitude and not exceeding 25\% in any magnitude bin brighter than our 50\%-completeness limit $m_{i'} \approx$ 23.5 (black solid line).}
\label{contam fig}
\end{figure}

%% file: fig_completeness.tex
\begin{figure}[!tbp]
\includegraphics[scale=0.575,angle=0]{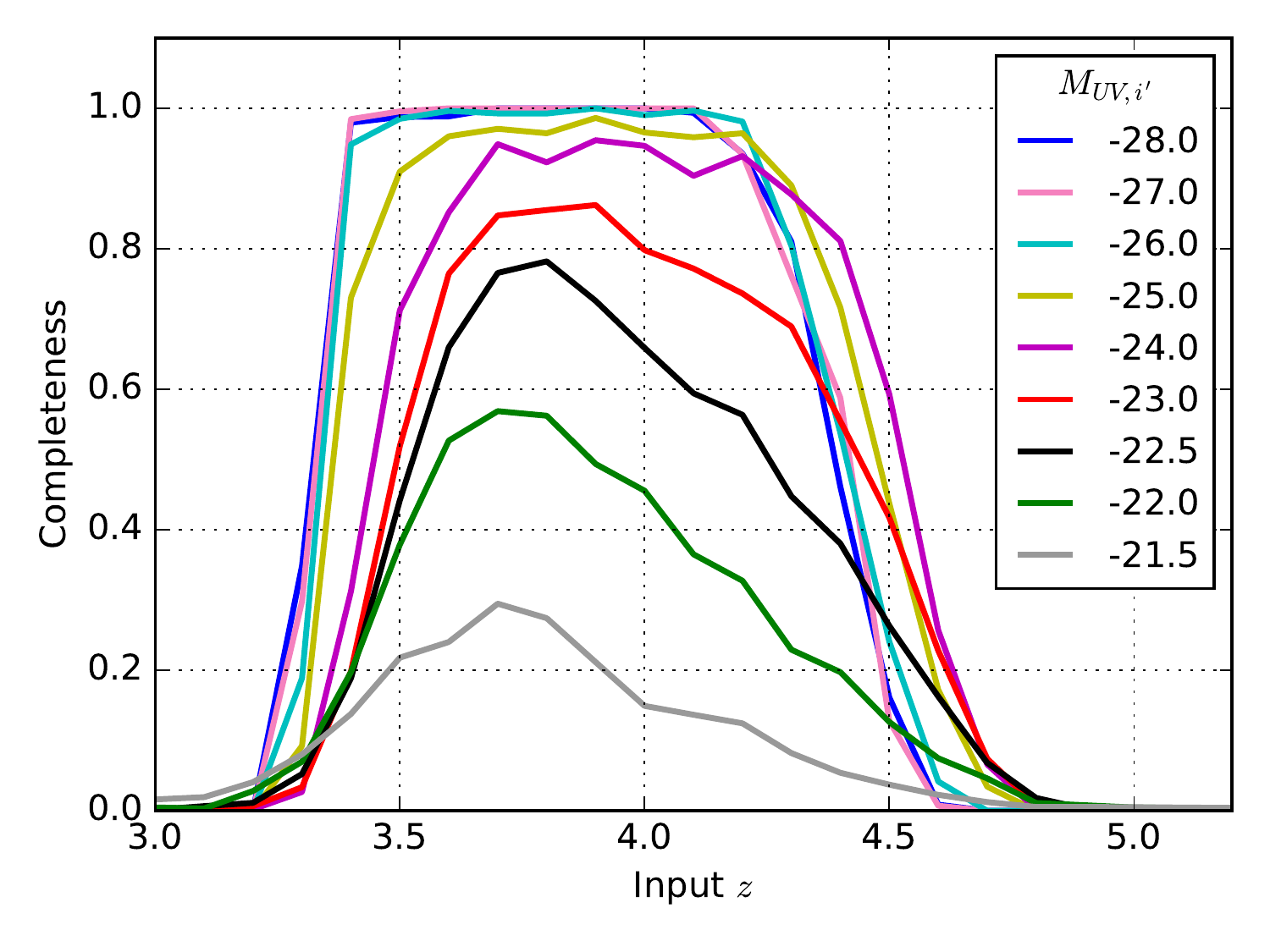}
\caption{The results of our completeness simulations, showing the fraction of simulated sources recovered as a function of input redshift. Each colored line is for the corresponding 0.5 magnitude bin according to the legend. We define the 50\%-completeness limit as the absolute magnitude curve with an integral value of less than 50\% the area under the average of the $M_{1450}=-25$ to $M_{1450}=-28$ curves. We find the 50\%-completeness to be $M_{1450}=-22.0$. }
\label{completeness fig}
\end{figure}

%% file: s4_results.tex
\section{Results} \label{results}

\subsection{The Rest-Frame UV z=4 Luminosity Function}
We utilize the effective volume method to correct for incompleteness in deriving our luminosity function. The effective volume ($V_{eff}$) can be estimated as 
\begin{equation}
V_{eff}(M_{i'})=\int\frac{dV_C}{dz}C(M_{UV,i'},z)dz,
\label{V_eff}
\end{equation}
\noindent where $\frac{dV_C}{dz}$ is the co-moving volume element, which depends on the adopted cosmology, and $C(M_{UV,i'},z)$ is the completeness as calculated in Section \ref{completeness section}. The integral was evaluated over $z=3-5$.

To calculate the luminosity function, we convert the apparent $i$-band AB magnitudes ($m_{i'}$) to the absolute magnitude at rest-frame 1500 \AA \ ($M_{UV,i'}$) using the following formula
\begin{equation}
M_{UV,i'}= m_i-5\log(d_L/10 \text{pc})+2.5\log(1+z),
\label{M_UV}
\end{equation}
\noindent where $d_L$ is the luminosity distance in pc. The second and third terms of the right side are the distance modulus.

\input{fig_our_data.tex}

Our UV luminosity function is shown as red diamonds in Figure \ref{our_data}. We note that we do not include the luminosity function data points in bins $M_{UV,i'}\geq$-22 in our analysis as these bins are below our 50\%-completeness limit as discussed in Section \ref{completeness section}, where the completeness corrections are unreliable due to the low S/N of our data. This is confirmed by comparing to the \textit{HST} CANDELS results in these same magnitude bins, which are more reliable due to their higher S/N. In Figure \ref{our_data} we also include the UV luminosity function of $z=4$ UV-selected galaxies from deeper \textit{Hubble} imagining by \cite{fink15} as green squares. In the bin where we overlap with this dataset ($M_{UV,i'}=$ -22.5) both results are consistent, however, if the luminosity function derived from Hubble imaging is extrapolated to brighter magnitudes, it would fall off more steeply than our luminosity function. While our luminosity function declines from $M_{UV,i'}=-$22 to $M_{UV,i'}=-$24, it flattens out at brighter magnitudes before turning over again. 

As our sample includes \emph{all} galaxies which exhibit a Lyman break, we expect that this flattening is due to the increasing importance of AGN at these magnitudes.  The large volume surveyed by SDSS data has led to the selection and spectroscopic follow-up of AGNs at many redshifts. SDSS AGN studies have found the AGN UV luminosity function to exhibit a double power law (DPL) shape \citep[e.g.,][]{richards06}.  In Figure \ref{our_data} we show as red "x"s the $z=4$ AGN UV number densities derived from the SDSS DR7 catalog \citep{schneider10} by \citet{akiyama17}, who select AGNs to $M_{UV,i'}>-28.9$. We can see that at the magnitudes where we overlap $-27\lesssim M_{UV,i'}\lesssim-26$, the agreement is excellent with our data. The only difference is at $M=-$28, where our survey detects two quasars when the AGN luminosity function by \citet{akiyama17} would predict less than one in our volume. We attribute this difference to cosmic variance. By combining our data with the star-forming galaxy number densities from Hubble imaging and the SDSS AGN number densities, our data can potentially provide a robust measurement of the bright-end slope of the star-forming galaxy luminosity function \textit{and} the faint-end slope of the AGN luminosity function, which we explore in the following section.

\subsection{Fitting the Luminosity Function} \label{fitting lf}
With our luminosity function in agreement with the faint end of the AGN luminosity function from SDSS DR7 and the bright end of the star-forming galaxy luminosity function from CANDELS, we attempt to simultaneously fit empirically motivated functions to both components. For the AGN component we use a DPL function motivated by the AGN UV luminosity function work on large, homogeneous quasar samples \citep[e.g.,][and references therein]{boyle00, richards06, croom09, hopkins07}. The function form of a DPL follows

\begin{equation}
\Phi(M) = \frac{\Phi^{*}}{10^{0.4(\alpha+1)(M-M^{*})}+10^{0.4(\beta+1)(M-M^{*})}},
\label{dpl}
\end{equation}
\noindent where $\Phi^*$ is the overall normalization, $M^*$ is the characteristic magnitude, $\alpha$ is the faint-end slope, and $\beta$ is the bright-end slope.

For the star-forming galaxy UV luminosity function we consider sepraratley both a Schechter function and a DPL, as well as including magnification via gravitational lensing with both functions. The \citet{schechter76} function has been found to fit the star-forming galaxy UV luminosity function well across all redshifts \citep[e.g.,][]{steidel99, bouwens07, fink15}. The Schechter function is described as
\begin{equation}
\Phi(M) = \frac{0.4\ \ln(10)\  \Phi^{*}}{10^{0.4(\alpha+1)(M-M^{*})}e^{10^{-0.4(M-M^{*})}}},
\label{schechter}
\end{equation}
\noindent where $\Phi^*$ is the overall normalization, $M^*$ is the characteristic magnitude, and $\alpha$ is the faint-end slope.

We consider the effects of gravitational lensing on the shape of the star-forming galaxy UV luminosity function. Gravitational lensing can distort the shape and magnification of distant sources as the paths of photons from the source get slightly perturbed into the line of sight of the observer. \cite{lima10} showed that this magnification can contribute to a bright-end excess where the slope of the intrinsic luminosity function is sufficiently steep. A magnification distribution for a given source redshift must be estimated by tracing rays through a series of lens planes derived from simulations such as the Millennium Simulation as done by \cite{hilbert07}. \cite{van10} showed that a Schechter function corrected for magnification can fit the bright-end of the luminosity function at $z=3$ better than a Schechter fit alone. They inspect the sources that make up the excess and find nearby massive foreground galaxies or groups of galaxies that could act as lenses. We incorporate the effects of gravitational lensing in our fitting by creating a lensed Schechter function parameterization following the method of \citet{ono17} who adapts the method of \citet{wyithe11}. We also produce a lensed DPL function. After performing our simultaneous fitting method, which we describe in the following paragraph, we found there to be no difference in the best fitting parameters of the fits including and excluding the effects of lensing. This is consistent with the work of \citet{ono17} who found that taking into account the effects of lensing improves the galaxy luminosity function fit at $z > 4$ and not at $z = 4$ where a DPL fit is preferred. Therefore we do not consider the lensed parameterizations further.

\input{table_lf_fits.tex}

We employ a Markov Chain Monte Carlo (MCMC) method to define the posteriors on our luminosity function parameterizations. We do this using an IDL implementation of the affine-invariant sampler \citep{goodman10} to sample the posterior, which is similar in production to the \textit{emcee} package \citep{foreman-mackey13}. Each of the 500 walkers was initialized by choosing a starting position with parameters determined by-eye to exhibit a good fit, perturbed according to a normal distribution. We do not assume a prior for any of our free parameters.

We account for Eddington bias in our fitting routine. Rather than directly comparing the observations to a given model, we forward model the effects of Eddington bias into the luminosity function model, and compare this ``convolved" model to our observations.  We do this by, for each set of luminosity function parameters, realizing a mock sample of galaxies for that given function, where each galaxy has a magnitude according to the given luminosity function distribution.  The magnitude of each simulated object is perturbed by an amount drawn from a normal distribution centered on zero with a width equal to the real sample median uncertainty in the corresponding magnitude bin. After perturbing, we then re-bin the simulated luminosity distribution and this binned luminosity function is used to calculate the $\chi^2$ for that MCMC step. This is repeated for each step.

We ran our MCMC fitting algorithm twice. In both runs, we simultaneously fit a DPL function to the AGN component of the luminosity function while fitting the galaxy component. In the first run, we fit a Schechter function to the galaxy luminosity function, while in the second we fit a second DPL function to the galaxy luminosity function. We burn each chain for 2$\times$10$^{5}$ steps, which allows the chain to reach convergence for all free parameters, verified by examining the parameter distributions in independent groups of 10$^4$ steps, which cease to evolve much past 10$^5$ steps. We then measure the posterior for each parameter from the final 5$\times$10$^4$ steps. Our fiducial values for each parameter are then derived as the median and 68\% central confidence range from the posterior distributions. The best fitting parameters for the two fits and their corresponding $\chi^2$ values are listed in Table \ref{fit_table}. We over plot the fits to the luminosity function data in Figure \ref{our_data}. To calculate the $\chi^2$ for our fits, we must compare the observed data to the luminosity function fits after forward modeling the effects of Eddington bias into our luminosity function fits just as we did during the fitting process. The absolute residuals of our ``convolved" fits and the observed data are shown in the insert in Figure \ref{our_data}.

 These fitting results show that our data prefers the function that is a sum of two DPL functions, one for the AGN component and one for the galaxy component, henceforth the DPL+DPL Fit. The DPL+DPL Fit has a $\chi^2=42$ over the entire range considered, $-29<M_{UV,i'}<-17$. The fit that is a sum of a DPL AGN component and a Schechter galaxy component, henceforth the DPL+Schetcher Fit, has a $\chi^2=71$. We investigate which fit is preferred by the data using the Bayesian information criterion for the two fits \citep{liddle07}. The difference in the Bayesian information criterion value for our two fits can be defined as:

\begin{equation} \label{del_bic}
     \Delta\textrm{BIC} = \chi^{2}_{2} - \chi^{2}_{1} + (k_2 - k_1)\ \textrm{ln}\ N,
\end{equation} 
\noindent where $\chi^{2}_{2}$ is the $\chi^{2}$ for the DPL+Schechter Fit, $\chi^{2}_{2}$ is the $\chi^{2}$ for the DPL+DPL Fit, $k_2$ and $k_1$ are the number of fit parameters in the DPL+Schechter Fit and the DPL+DPL Fit, respectively, and $N$ is the number of data points used during fitting. We find a $\Delta\textrm{BIC}=26$ which suggests the DPL+DPL Fit is very strongly preferred over the DPL+Schechter Fit as $\Delta\textrm{BIC}$ exceeds a significance value of 10 \citep{liddle07}. Upon comparing the the fits' residuals we see the data points in the magnitude bins at $M_{UV,i'}=-24$ mag and $M_{UV,i'}=-23$ mag are driving the preference for the DPL+DPL Fit. If we exclude these two bins, the DPL+DPL Fit $\chi^{2}=41$, the DPL+Schechter Fit $\chi^{2}=58$, and the $\Delta\textrm{BIC}=14$ which indicates the DPL+DPL Fit is still strongly statistically preferred to the DPL+Schechter Fit. However, given that the difference between the fits is driven by just a few data points, we do not believe we can firmly rule out a Schechter form for the star-forming galaxy component.
\input{fig_agn_demos.tex}

\subsection{A Sample of Spectroscopically Confirmed AGNs} \label{agns}
Given our method of fitting the luminosity function with a component for AGNs, we explored the SDSS spectral catalog for spectroscopically confirmed AGNs in SHELA and considered the effectiveness of our selection procedure at recovering AGNs. 
This is crucial as our photometric redshift selection process did not include templates dominated by AGN emission, though the strong Lyman break inherent in these sources should still yield an accurate redshift.  To confirm this, we queried the spectral catalog from SDSS (DR13; \citealt{abareti17}) using the SDSS CasJobs website\footnote{http://skyserver.sdss.org/casjobs/} and cross-matched the results with our entire DECam catalog. Each match required a separation of less than 0.4" and the SDSS spectra to be unflagged (i.e., ZWARNING = 0). We found zero spectra with an SDSS classification of “Galaxy” and 53 classified as “AGN” with spectroscopic redshifts ($z_{\mathrm{spec}}$) greater than 3.2. The distribution of SDSS $z_{\mathrm{spec}}$ for this sample is shown in the left panel of Figure \ref{agn_demos} in blue. Of the 32 AGNs with $3.4<z_{\mathrm{spec}}<4.6$, 23 were selected by our $z_{\mathrm{phot}}$ selection process, with seven of the nine missed AGNs having $3.4<z_{\mathrm{spec}}<3.5$. The $z_{\mathrm{spec}}$ distribution for the AGN subsample selected by our selection procedure is also shown in the left panel of Figure \ref{agn_demos} in red. We used these samples to compute our differential completeness of AGNs and compared it to our simulated completeness in the right panel of Figure \ref{agn_demos}. We found our completeness of spectroscopically confirmed AGNs is consistent with our simulated completeness except in the $3.4<z<3.6$ bin where we recovered only two of the nine spectroscopically confirmed AGNs when our simulations predicted we should recover 7$-$8. This difference could be due to small number statistics. We investigated why the seven spectroscopically confirmed AGNs in the $3.4<z<3.6$ bin were not selected by our method and found that these sources had photometry, particularly the $u'$ and $g'$ bands, preferred by galaxies templates at lower redshift in our $z_{phot}$-fitting code EAZY. We attribute bright $u'$ and $g'$ fluxes to the larger far-UV continuum levels of AGNs or strong Lyman-$\alpha$ emission as compared to non-AGN galaxies, which would weaken the Lyman-$\alpha$ break in these sources. This could imply significant leaking Lyman-continuum radiation from these AGNs, which has implications on reionization \citep[e.g.,][]{smith16}. The reliability of our selection procedure to recover AGNs across the majority of our redshift range of interest and the fact that our luminosity function is consistent with the AGN luminosity function from SDSS suggests our incompleteness to AGNs is not substantial.

%% file: fig_our_data.tex
\begin{figure*}[!tbp]
\centering
\includegraphics[scale=0.8,angle=0]{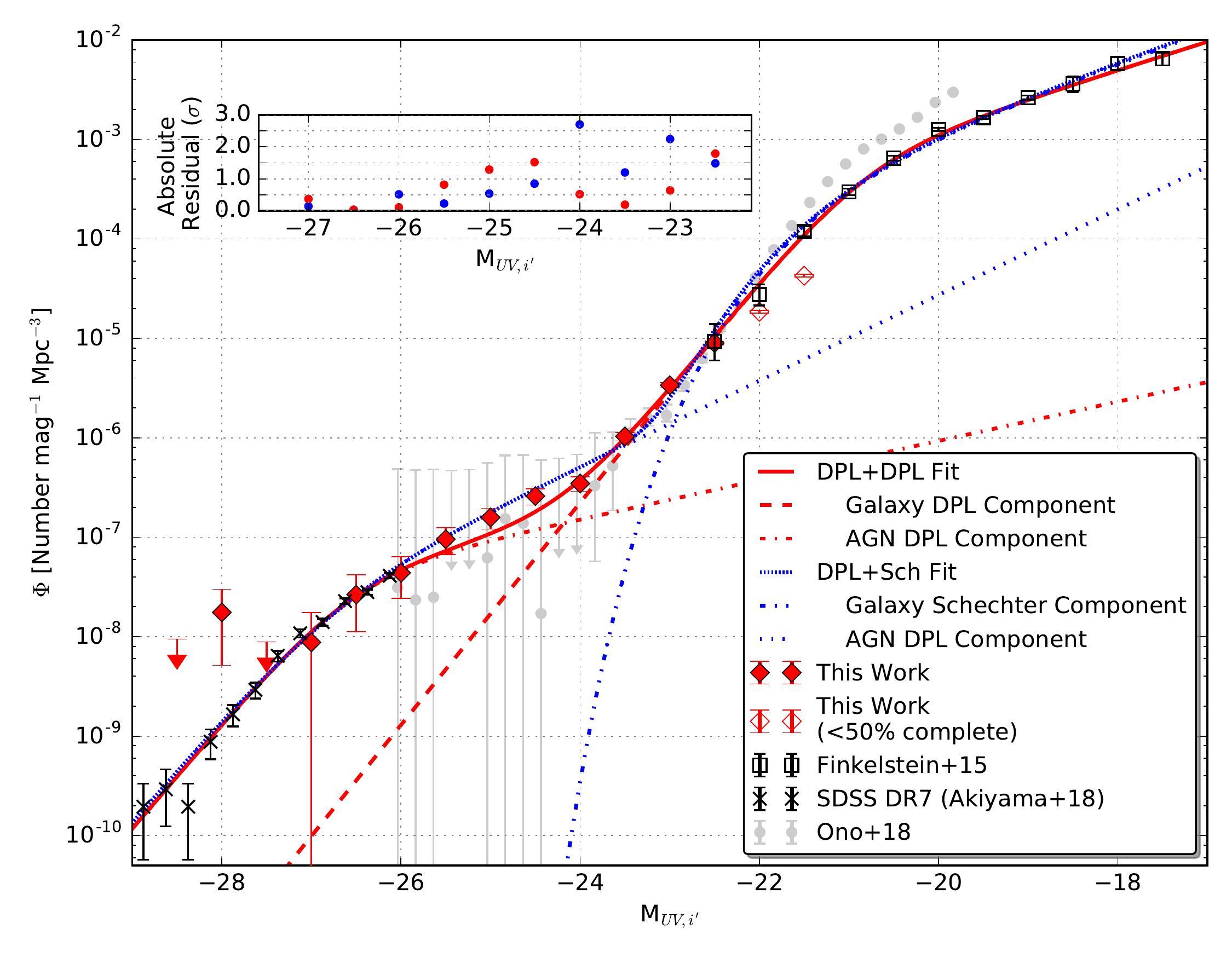}
\caption{The rest-frame UV $z=4$ luminosity function of star-forming galaxies and AGNs from the SHELA Field shown as red diamonds with Poisson statistic error bars. The open red diamonds are the luminosity function points in bins below our 50\%-completeness limit as discussed in Section \ref{completeness section}, where the completeness corrections are unreliable due to the low S/N of our data. We constrain the form of the luminosity function by including fainter galaxies from Hubble fields (\citealt{fink15}; open black squares) and brighter AGNs from SDSS DR7 (\citealt{akiyama17}; black ``x"s). For comparison, we overplot as gray circles the $g'$-band dropout luminosity function from the $>100$ sq. deg. HSC SSP by \citet{ono17}, which shows lower number densities and larger error bars in the regime ($M_{UV,i'}<23.5$ mag) where AGN likely dominate (see Section \ref{intro}). Our measured luminosity function is consistent with these works where they overlap. Our two best-fitting functional forms are shown, as discussed in Section \ref{fitting lf}. The fit with the smallest $\chi^2$ is the sum of two DPL functions (DPL+DPL Fit; red solid line), one for the AGN component (red dash-dotted line) and one for the galaxy component (red dashed line). The second best fit (DPL+Sch Fit; blue densely dotted line) is comprised of a DPL function for the AGN component (blue dotted line) and a Schechter function for the galaxy component (blue dash-dot-dotted line). The absolute value of the residuals of the two fits are shown in the inset plot for a subset of the data in units of the uncertainty in each bin. The data favors the DPL+DPL Fit suggesting the $M_{UV,i'}=-23.5$ mag bin is dominated by star-forming galaxies, though this is dominated by the observed number densities in just a few bins, thus it is difficult to rule out a Schechter form for the galaxy component.}
\label{our_data}
\end{figure*}

%% file: table_lf_fits.tex
\begin{deluxetable*}{ccccccccccc}[ht]
\tablecolumns{11} 
\tablecaption{Fit Parameters for Luminosity Functions\label{fit_table}}
\tablehead{
\colhead{}    &  \multicolumn{4}{c}{AGN Component} &   \colhead{}   & 
\multicolumn{4}{c}{Galaxy Component} &   \colhead{}\\ 
\cline{2-5} \cline{7-10} \\ 
\colhead{Fit Name} & \colhead{Log $\Phi^{*}$}   & \colhead{$M^{*}$}    & \colhead{$\alpha$} & \colhead{$\beta$} & 
\colhead{}        & \colhead{Log $\Phi^{*}$}   & \colhead{$M^{*}$}    & \colhead{$\alpha$} & \colhead{$\beta$} & \colhead{$\chi^2$} \\
\colhead{} & \colhead{}   & \colhead{(mag)}    & \colhead{[Faint End]} & \colhead{[Bright End]} & 
\colhead{}        & \colhead{}   & \colhead{(mag)}    & \colhead{[Faint End]} & \colhead{[Bright End]} & \colhead{}} %
\startdata 
DPL+DPL &
-7.32$^{+0.21}_{-0.18}$\phn &
-26.5$^{+0.4}_{-0.3}$\phn &
-1.49$^{+0.30}_{-0.21}$\phn &
-3.65$^{+0.21}_{-0.24}$\phn & 
& 
-3.12$^{+0.09}_{-0.10}$\phn &
-20.8$^{+0.16}_{-0.15}$\phn &
-1.71$\pm$0.08\phn &
-3.80$\pm$0.10\phn &
42\\
DPL+Sch &
-7.48$^{+0.58}_{-0.34}$\phn &
-26.7$^{+1.1}_{-0.4}$\phn & 
-2.08$^{+0.18}_{-0.11}$\phn &
-3.66$^{+0.68}_{-0.34}$\phn & 
&
-3.25$\pm$0.06\phn &
-21.3$\pm$0.06\phn &
-1.81$\pm$0.05\phn &
\nodata &
71 \\ %
\enddata 
\tablecomments{$\Phi^{*}$ in units of Mpc$^{-3}$ mag$^{-1}$. The parameters for the galaxy component of DPL+Sch correspond to a Schechter function (Equation \ref{schechter}) and the remaining sets of parameters correspond to a double power-law function (Equation \ref{dpl}).}  
\end{deluxetable*}

%% file: fig_agn_demos.tex
\begin{figure}[!tbp]
\includegraphics[scale=0.6,angle=0]{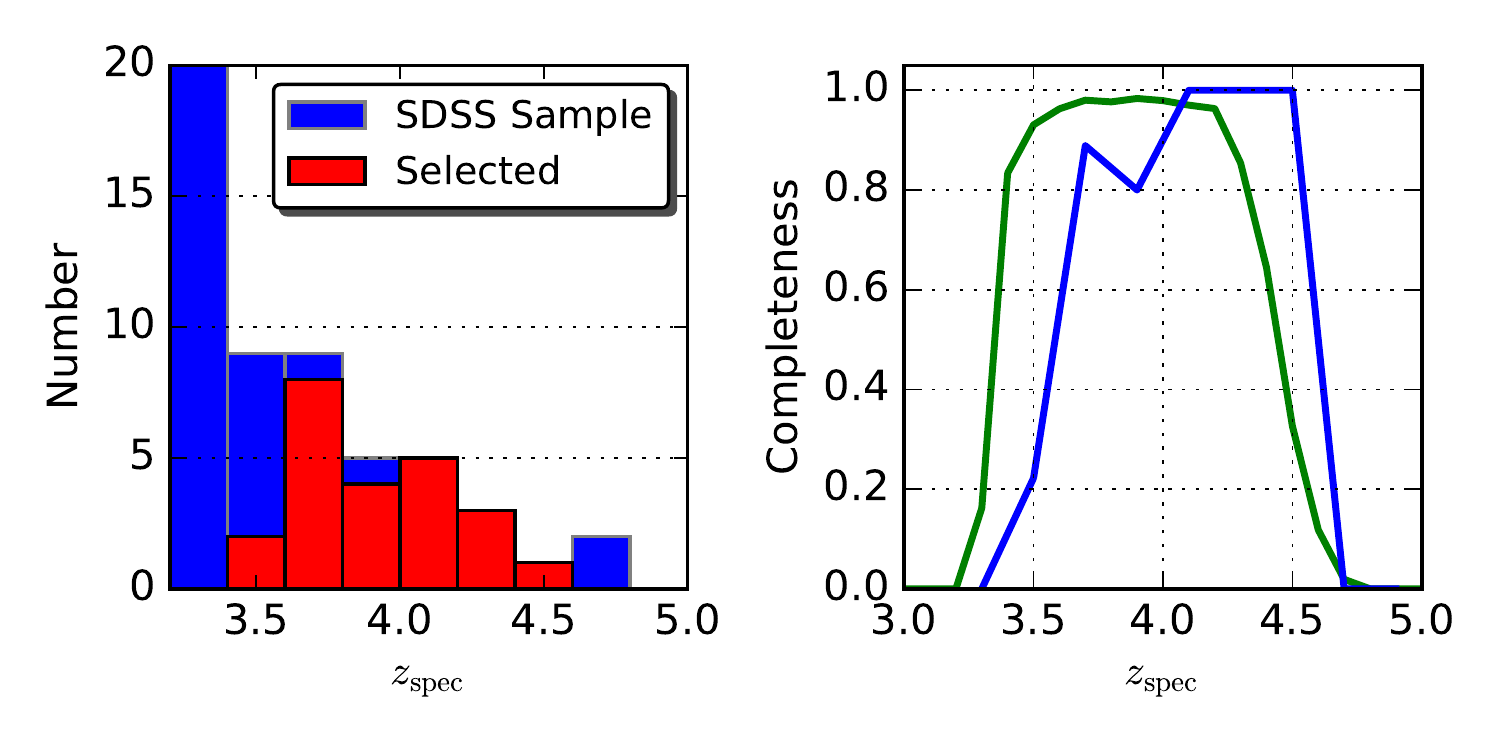}
\caption{Left: The SDSS spectroscopic redshift distribution of AGNs in SHELA (blue) overplotted with the same distribution for the sub-sample selected by our selection procedure (red). Right: The differential completeness fraction of AGNs with SDSS spectroscopic redshifts (blue) and the expected differential completeness fraction for objects with a comparable $M_{UV,i'}$ (green). Our completeness to AGNs is similar to what we expect from our simulations of galaxies except at $z=3.5$ where we are less complete to AGNs due to significant $u'$ and $g'$ flux driving the redshift probability distribution functions to peak at redshifts below our selection window (see Section \ref{agns} for details).}
\label{agn_demos}
\end{figure}

%% file: s5_discussion.tex
\section{Discussion} \label{discussion}

\subsection{Comparison to z=4 Galaxy Studies}

We compare our derived fits to the star-forming galaxy luminosity function to measurements from the literature in Figure \ref{galaxy_data}. At magnitudes fainter than $M_{UV,i'}>-22$, the DPL galaxy component of the data-preferred DPL+DPL Fit is very similar to the Schechter component of the DPL+Schechter Fit. These results are in strong agreement with the luminosity function from the CFHT Deep Legacy Survey by \citet{van10} at all magnitudes where they overlap. They are also in strong agreement with the luminosity function from \textit{Hubble} legacy survey data by \citet{fink15} which is included in our fitting. The luminosity function from the \textit{Hubble} legacy survey data by \citet{bouwens15} and the luminosity function from \citet{ono17} using Subaru Hyper Suprime-Cam (HSC) data across 100 deg$^{2}$ in the HSC Subaru strategic program (SSP) are generally consistent with the results presented here. The HSC SSP luminosity function and that from \citet{bouwens15} have volume densities of $\sim$2x larger than the work by \citet{van10} and \citet{fink15} at magnitudes fainter than $M_{UV,i'}>-21$. This factor is larger than the $\sim$10\% uncertainty expected due to cosmic variance in the \textit{Hubble} fields, which cover ~50x less area than the HSC SSP \citep{fink15}. We do not know the cause of this difference, though we note that the HSC SSP selection was done with optical imaging only, and we found in our study that without the addition of the IRAC data, the contamination rate was significantly higher, although we acknowledge there are multiple differences in the selection techniques between these studies.

At the bright end, the luminosity function from \citet{ono17} extends to magnitudes brighter than $M_{UV,i'}<-22.5$ and falls between the bright end of our DPL galaxy component and our Schechter galaxy component. \citet{ono17} find their luminosity function is best fit by a DPL with a steeper bright-end slope ($\beta=-4.33$) than our galaxy DPL component ($\beta=-3.80$). We also include the $z=4$ galaxy luminosity function from the 4 deg$^2$ ALHAMBRA survey by \citet{viironen17} who used a $z_{\mathrm{phot}}$ PDF analysis to create a luminosity function marginalizing over both redshift and magnitude uncertainties. \citet{viironen17} find volume densities at the bright-end larger than existing luminosity functions. In fact, their luminosity function follows a Schechter function with a normalization offset of $\sim$0.5 dex above the Schechter form from \citet{fink15}. The cause of this difference is unclear.

\subsubsection{Comparison to Semi-Analytic Models}
Figure \ref{galaxy_data_sams} shows the predicted $z=4$ UV luminosity functions from \citet{yung18}.  These predictions come from a set of semi-analytic models (SAMs), which contain the same physical processes as the models presented in \citet{somerville15}, but have been updated and recalibrated to the Planck 2016 Cosmological parameters. We note that while these models include black hole accretion and the effects of AGN feedback, we do not examine the contribution to the UV luminosity function from black hole accretion here, and focus instead on the contribution from star formation. Their fiducial model with dust (solid black line) assumes that the molecular gas depletion time is shorter at higher gas densities (as motivated by observations), leading to an effective
redshift dependence as high redshift galaxies are more compact and have denser gas on average. On a SFR surface density versus gas surface density plot, this model would show a steeper dependence of star formation rate density on gas surface density than the classical Kennicutt-Schmidt relationship \citep[e.g.,][]{kennicutt12}, above a critical gas surface density \citep[for details see][]{somerville15}.  We also show their model with a fixed molecular gas depletion time, similar to that used in \citet{somerville15}, as seen in local spiral galaxies \citep[e.g.,][]{bigiel08, leroy08}.

The \citet{yung18} dust models assume that the V-band dust optical depth is proportional to the cold gas metallicity times the cold gas surface density. The UV attenuation is then computed using a Calzetti attenuation curve and a ``slab" model \citep[for details see][]{somerville15}. The normalization of the dust optical depth (physically equivalent to the dust-to-metal ratio) is allowed to vary as a function of redshift, and was adjusted to fit the bright end of the luminosity from the previously published compilation of UV luminosity observations from \citet{fink16}.  (It was not adjusted to fit the new observations presented here).

\input{fig_galaxy_data.tex}
\input{fig_galaxy_data_sams.tex}

At the faint end, the model predictions are insensitive to the assumed star formation efficiency, and mainly
reflect the treatment of outflows driven by stars and supernovae. The models have a higher normalization than the observed luminosity function (which at these magnitudes comes from the CANDELS dataset), although the faint-end slope is similar.  This could be caused by stellar feedback in the models being too weak, as these models were tuned to match the $z=0$ stellar mass function \citep{somerville15}.  This suggests that stellar feedback is stronger/more efficient at $z=4$ (i.e., mass loading rates are higher, or re-infall of ejected gas is slower) than at $z=0$ \citep[also see][]{white15}.  This was also seen by \citet{song16} when comparing the $z=4$ stellar mass function to a number of similar models -- the observed stellar mass function also had a lower normalization at $z=4$; as the stellar mass function steepened from $z=4$ to 8, this discrepancy weakened, hence their conclusion of a weakening impact of feedback on the faint end with increasing redshift.

At the bright end, the \citet{yung18} model with dust is consistent with both of our fits to $M_{UV,i'} > -$22.5, lying closer to the Schechter fit at brighter magnitudes.  Interestingly, at these bright magnitudes both the Schechter and DPL fit rule out at high significance the model with dust and fixed molecular gas depletion time, indicating that star formation must scale non-linearly with molecular gas surface density (or some related quantity which evolves with redshift).  This implies that star formation is more efficient at $z=4$ than today.  This is of course dependant on the dust model in these simulations -- if bright galaxies had no dust, then these model predictions (which include dust; models without dust are shown for comparison) may not be accurate.  However, there is a variety of evidence that bright/massive UV-selected galaxies at these redshifts do contain non-negligible amounts of dust \citep[e.g.,][]{finkelstein12,bouwens14}, and there is a not-insignificant population of extremely massive dusty star-forming galaxies already in place \citep[e.g.,][]{casey14}.  Finally, \citet{fink15} compared a similar set of models to the CANDELS-only UV luminosity functions, finding that models with a diffuse dust component only (e.g., no birth cloud) provided the best match to the data (albeit at fainter magnitudes than considered here).

\subsection{Comparison to $z=4$ AGN Studies}

We compare our derived AGN luminosity function fits to measurements from the literature in Figure \ref{agn_data}. At the bright end, our fits are consistent with previous studies where they overlap ($-27.5<M_{UV,i'}<-25.5$). The previous studies include a study by \citet{richards06} who used the $z=4$ AGN SDSS DR3 sample and a study by \citet{akiyama17} who selected $z=4$ AGNs from SDSS DR7. We convert the magnitudes $M_{i(z=2)}$ at $z=3.75$ from \citet{richards06} to $M_{UV,i'}$ at $z=3.8$ by adding an offset of 1.486 mag \citep{richards06}.

\input{fig_agn_data.tex}

At the faint end, we compare our fits to results from studies that derive AGN luminosity functions from spectroscopic observations of candidates selected via color and size criteria (\citealt{glikman11}, \citealt{ikeda11}, and \citealt{niida16}). In addition, we include studies that rely on a $z_{\mathrm{phot}}$ selection using deep multi wavelength photometry in the COSMOS field \citep{masters12} and the CANDELS GOODS-S field \citep{giallongo15} where \citet{giallongo15} had the additional criteria of requiring an an X-ray detection of $F_X > 1.5 \times 10^{-17} \mathrm{erg\ cm}^{-2}\ \mathrm{s}^{-1}$ in deep 0.5-2 keV \textit{Chandra} imaging. The average redshift of these samples is $z=4$, slightly higher than our sample. For a consistent comparison with other works we scale the luminosity functions of \citet{glikman11}, \citet{ikeda11}, \citet{niida16}, and \citet{masters12} up by a factor of 1.3 using the redshift evolution function $(1+z)^{-6.9}$ from \citet{richards06}. The sample used by \citep{giallongo15} had a redshift range of $4<z<4.5$, so we the scale the luminosity function by a factor of 1.8. Finally, we consider the luminosity function of \citet{akiyama17} derived from the Hyper Suprime-Cam Wide Survey optical photometry. 

The faint end of the AGN DPL component in our DPL+DPL Fit predicts volume densities at $-26<M_{UV,i'}<-23.5$ about 0.3 dex lower than those found by \citet{richards06}, \citet{ikeda11}, \citet{masters12}, \citet{niida16}, \citet{akiyama17}. However, the luminosity function of \citet{akiyama17} flattens and falls towards our fit at $M_{UV,i'}\sim-22$. Our faint-end slope ($\alpha=-1.49^{+0.30}_{-0.21}$) is in agreement with the values found by these studies. The cause of the 0.3 dex difference is unclear.

In the case of the AGN DPL component in our DPL+Schechter Fit, the steeper faint-end slope ($\alpha=-2.08^{+0.18}_{-0.11}$) predicts volume densities in agreement with \citet{glikman11} and \citet{giallongo15}, who predict a significantly higher volume density of faint AGNs than the other studies. 
We note that while these studies have small numbers of AGN per magnitude bin, which may make the samples more sensitive to cosmic variance and false positives, \citet{glikman11} selects candidates from a relatively large survey area of 3.76 deg$^2$ and observes broad emission lines, indicative of quasars, in every candidate spectra. Furthermore, \citet{giallongo15} requires candidates to have a significant X-ray detection in \textit{Chandra} imaging. 

In summary, our observed $z=4$ UV luminosity function is best fit by the DPL+DPL Fit, but both the DPL+DPL and the DPL+Schechter fits show agreement with existing AGN luminosity function studies at the bright end and around the knee. At the faint end the DPL+DPL predicts smaller volume densities of AGNs than other studies while the DPL+Schechter predicts among the largest volume densities at the faintest magnitudes. %

\subsection{Comparing Predictions of Our Two Fits - Is the UV luminosity function a double power law?}
Our two fits differ in two ways: 1.) The functional form used to fit the star-forming galaxy component is a DPL in the DPL+DPL Fit and a Schechter in the DPL+Schechter Fit, and 2.) The component that accounts for the excess over an extrapolated Schechter at the bright end of existing star-forming galaxy luminosity functions. The DPL+Schechter accounts for the excess with a steeper faint end of the DPL AGN component while the DPL+DPL Fit accounts for the majority of the excess with the bright end of the DPL star-forming galaxy component. Thus the fits predict dramatically different compositions of sources making up the bin ($M_{UV,i'}=-23.5$ mag) at the center of the excess. In this bin the DPL+Schechter Fit predicts AGNs outnumber galaxies by a 17:1 ratio (an AGN fraction of $\sim$94\%), while the DPL+DPL Fit predicts galaxies to outnumber AGNs by a 4.3:1 ratio (an AGN fraction of $\sim$18\%). A simple experiment to test these predictions would be to use ground-based optical spectroscopy to follow-up a fraction of the 298 high-$z$ candidates we find in the $M_{UV,i'}=-23.5$ bin and count the fraction of spectra exhibiting broad emission lines and/or highly ionized lines (e.g., N V, He II, C IV, Ne V) indicating accretion onto supermassive black holes. This experiment would also provide an independent measurement of our contamination fraction which we estimate via simulations (Section \ref{est contam}). The measured fraction of AGNs can provide strong empirical proof in favor of the DPL+DPL Fit or the DPL+Schechter Fit.

One drawback of fitting the total UV luminosity function is the unknown contribution of composite objects. The total UV luminosity function likely contains a population of composite objects with both AGN activity and star-formation contributing to their observed UV flux. This population may cause functions like a DPL or Schechter function, which have been used to fit AGN-only and star-forming-galaxy-only luminosity functions, to be poor fits to the total UV luminosity function. Spectroscopic follow-up as described in this section may aid in elucidating the frequency and impact of composite objects.

\input{table_uv_output.tex}
\subsection{Rest-Frame UV Emissivities}

Here we calculate the rest-frame UV emissivities (also known as specific luminosity densities) and compare the output from AGNs and galaxies. We calculate the rest-frame UV emissitvity of AGNs and galaxies by integrating each luminosity function from  $-30 < M_{UV,i'} < -17$. Results are shown in Table \ref{uv_table}. In the case of DPL+DPL where the galaxy component of the UV luminosity function is represented by a DPL function and the AGN component is represented by a DPL, galaxies produce a UV luminosity density at 1500 \AA\ ($\rho_{1500}$) greater than that of AGNs by a factor of $\sim$90. In the case of the DPL+Schechter Fit where the galaxy component of the UV luminosity function is instead represented by a DPL function, the galaxy $\rho_{1500}$ is greater than the AGN $\rho_{1500}$ by a factor of $\sim$19. In either case, galaxies are the dominant non-ionizing UV emitting population, though if AGNs do have a steeper faint-end slope (as would need to be the case if galaxies follow a Schechter form), then AGNs are non-negligible.

\input{fig_uv_lum_den.tex}

We convert the AGN $\rho_{1500}$ to a UV luminosity density at 912 \AA\ ($\rho_{912}$) and compare our results to other studies by reproducing Figure 1 from \citet*{madau15} in Figure \ref{uv_lum_den}. To convert $\rho_{1500}$ to $\rho_{912}$ we assume an AGN spectral shape of a DPL with a slope of
$\alpha_{\nu}=-1.41$ \citep{Shull:2012} between 1000 \AA\ and 1500 \AA\ and a slope of $\alpha_{\nu}=-0.83$ \citep{stevans:2014} between 912 \AA\ and 1000 \AA\ . We assume an AGN ionizing escape fraction of unity as is found for most bright AGNs \citep{giallongo15}. Results are shown in Table \ref{uv_table}. If we directly follow the work of \citet{giallongo15} and assume a spectral shape with $\alpha_{\nu}=-1.57$ \citep{Telfer:2002} between 1200 \AA\ and 1500 \AA\ and a slope of $\alpha_{\nu}=-0.44$ \citep{VandenBerk:2001} between 912 \AA\ and 1200 \AA\ , we find $\rho_{912}$ values that are 10\% smaller. As shown in Figure \ref{uv_lum_den}, our two fits predict values of $\rho_{912}$ straddle existing values from observations. The $\rho_{912}$ predicted by our preferred DPL+DPL Fit (upward-pointed red triangle) falls below the line by \citet{hopkins07} and is too small to keep the IGM ionized at $z\sim4$. On the other hand, the DPL+Sch Fit predicts a $\rho_{912}$ (downward-pointed red triangle) that falls near the points from studies that found numerous faint AGNs at $z\sim4$ \citealp{glikman11, giallongo15}. This would imply the AGN population could alone contribute enough hydrogen ionizing emission required to keep the universe ionized at $z\sim4$ and suggests AGNs may have played a significant role at early times.  This scenario can be further constrained by the spectroscopic experiment proposed in the preceding sub-section.

%% file: fig_galaxy_data.tex
\begin{figure}[!tbp]
\centering
\includegraphics[scale=0.6,angle=0]{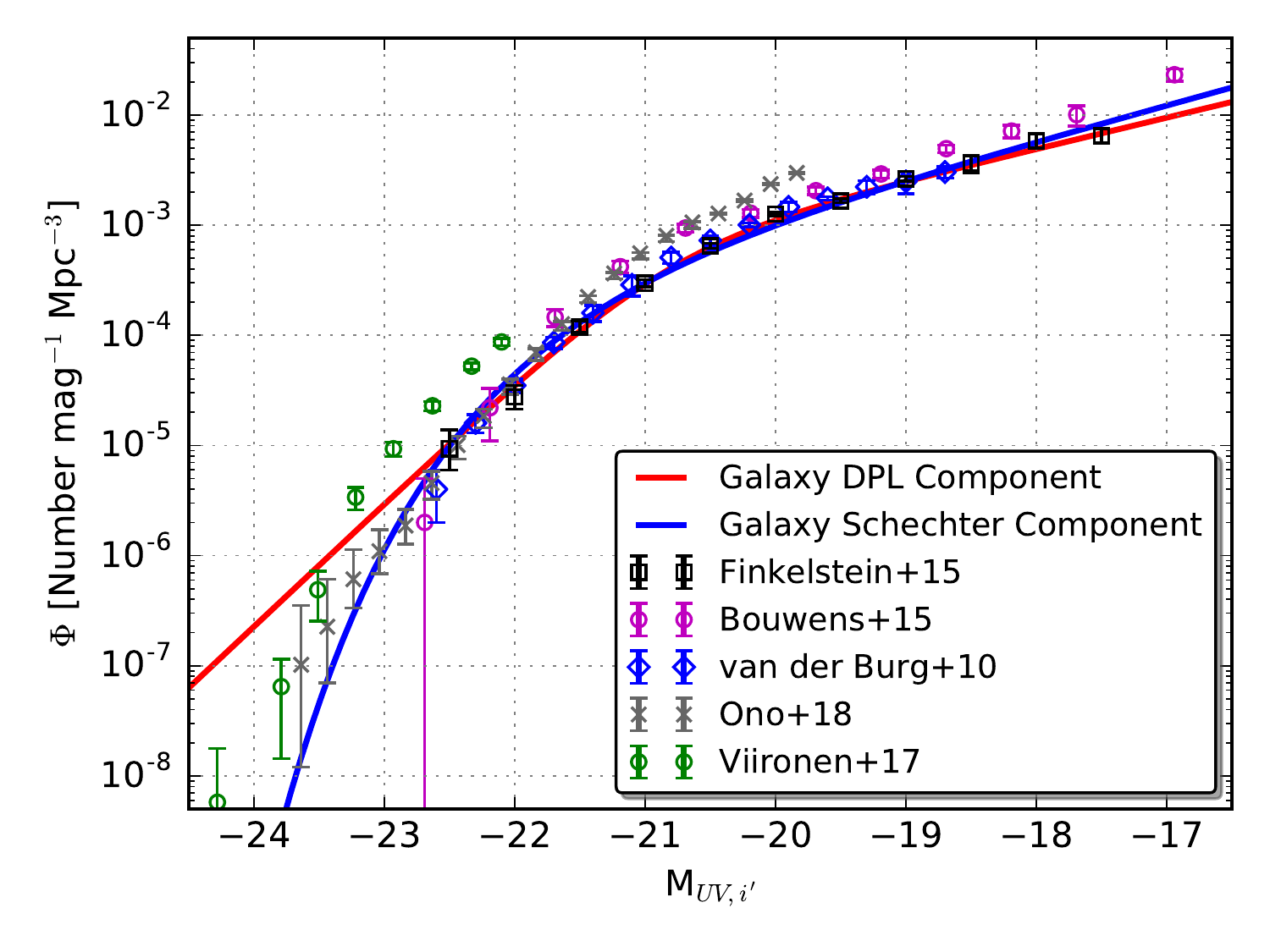}
\caption{The star-forming galaxies components of our fits to the total $z=4$ rest-frame UV luminosity function compared to the data from other star-forming galaxy studies See the legend insert for the list of compared works. Our DPL galaxy luminosity function is in agreement with luminosity functions from the literature around the knee and to fainter magnitudes, but has the shallowest bright-end slope ($\beta=-3.80$).}
\label{galaxy_data}
\end{figure}

%% file: fig_galaxy_data_sams.tex
\begin{figure}[!tbp]
\includegraphics[scale=0.6,angle=0]{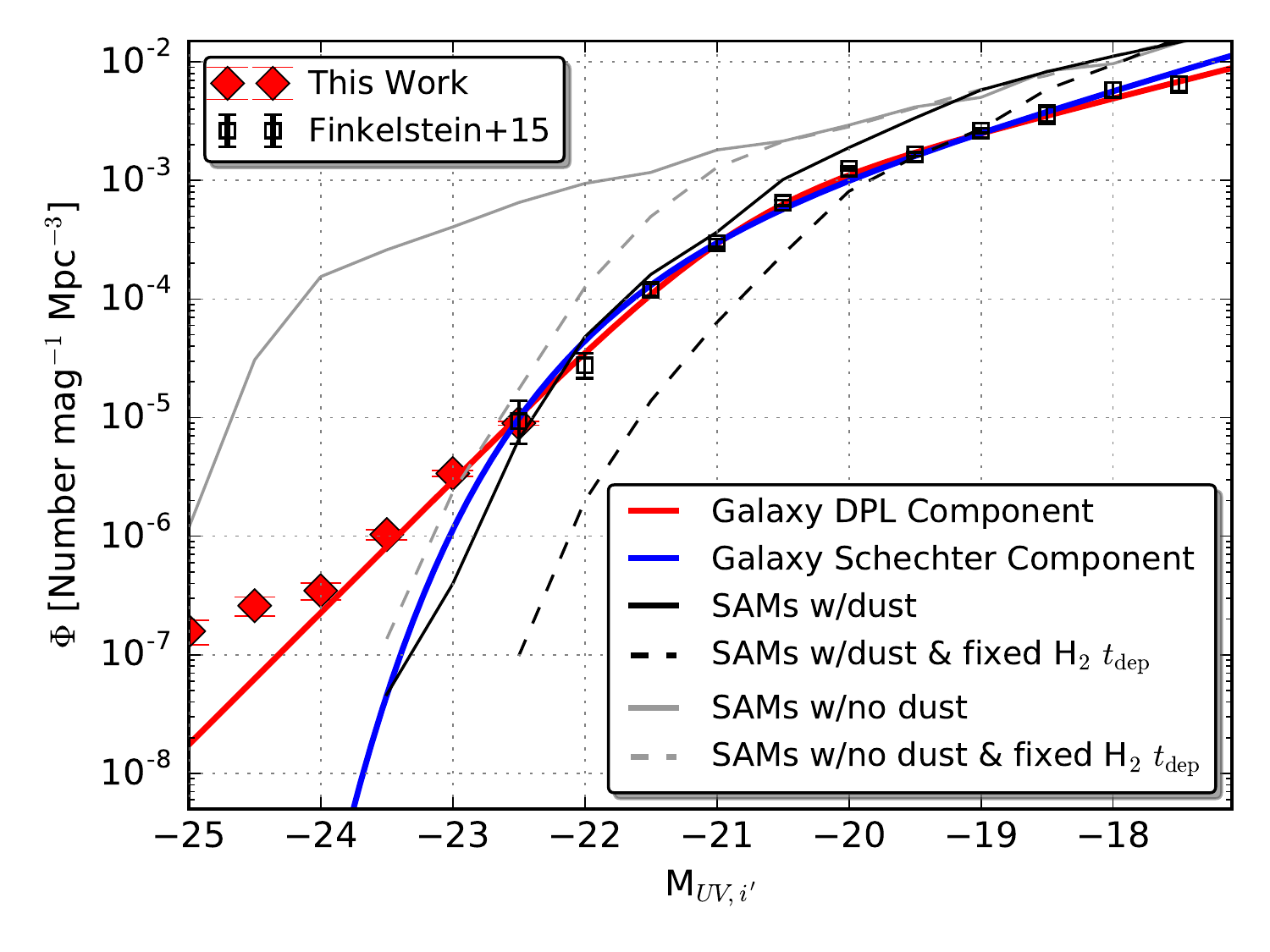}
\caption{The star-forming galaxy component of our fits to the total $z=4$ rest-frame UV luminosity function compared to predictions from SAMs \citep{yung18}. Our measured total luminosity function (including star-forming galaxies and AGNs) is shown as red diamonds with Poisson statistic error bars. The luminosity function from \citet{fink15} is shown as open black squares. At the faint end the luminosity functions from the SAMs with an evolving H$_2$ gas depletion time (solid black line) and a fixed H$_2$ gas depletion time (dashed black line) have a higher normalization than the observed luminosity function suggesting stellar feedback in the models is too weak. At the bright end the model with a fixed molecular depletion timescale is robustly ruled out by either of our parameterized fits, suggesting that star formation scales with molecular gas surface density, and thus is more efficient at $z=4$ than today.}
\label{galaxy_data_sams}
\end{figure}

%% file: fig_agn_data.tex
\begin{figure*}[!tbp]
\centering
\includegraphics[scale=0.6,angle=0]{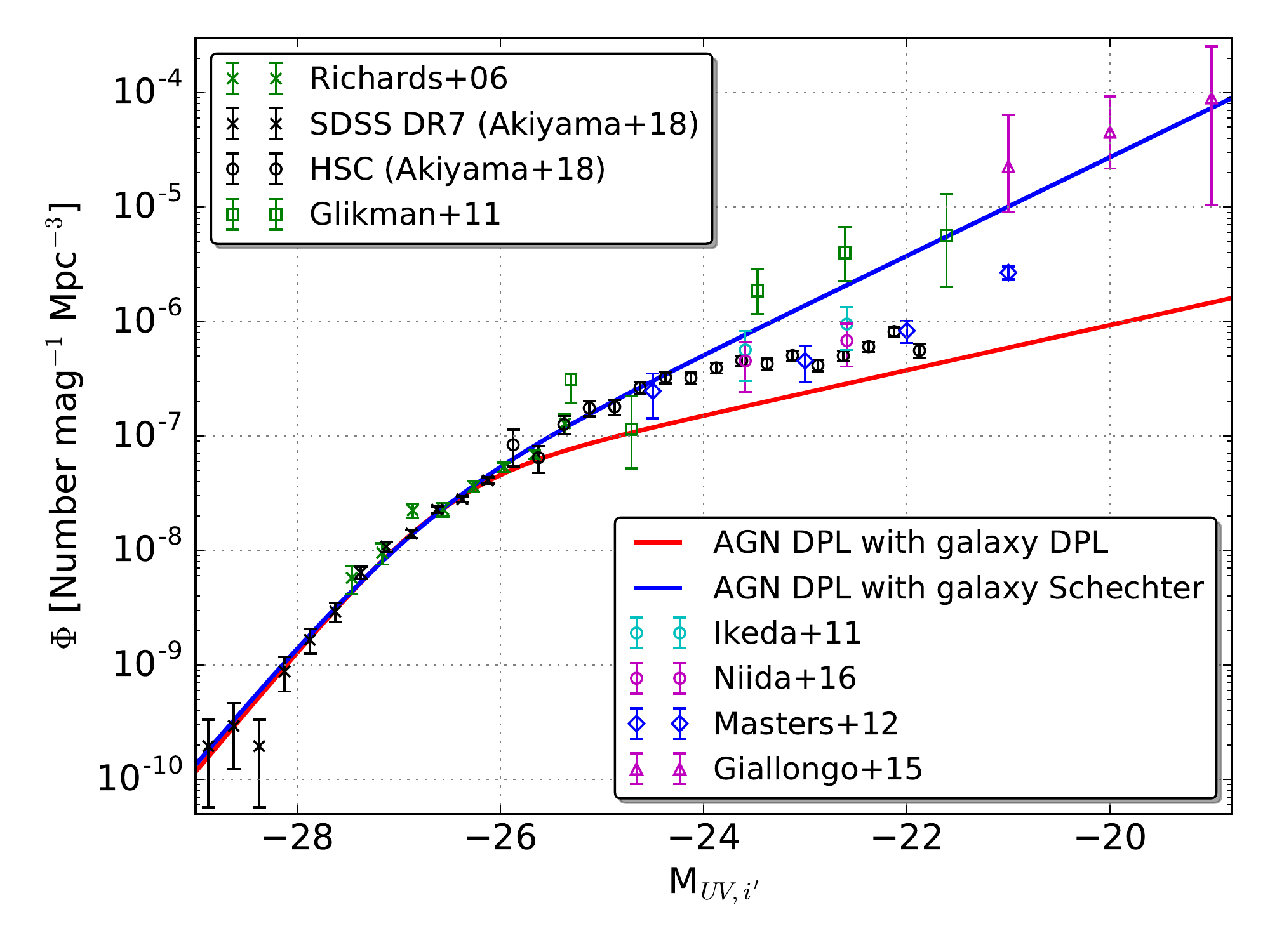}
\caption{The AGN components of our fits to the total $z=4$ rest-frame UV  luminosity function compared to the data from other AGN studies. See the legend insert for the list of compared works. The AGN luminosity function from the DPL+DPL Fit has number densities similar to existing luminosity function measurements at $M_{UV,i'}<-25$ and predicts relatively low number densities at fainter magnitudes. The DPL+Schechter Fit has an AGN faint-end slope that predicts number densities at the larger end of the range previously published. }
\label{agn_data}
\end{figure*}

%% file: table_uv_output.tex
\begin{deluxetable}{cccccc}[htb!]
\tablecaption{UV Luminosity Densities\label{uv_table}}
\tablecolumns{6} 
\tablehead{
\colhead{}    &  \multicolumn{2}{c}{AGN Component} &   \colhead{}   & 
\multicolumn{2}{c}{Galaxy Component} \\ 
\cline{2-3} \cline{5-6} \\ 
\colhead{Fit Name} & \colhead{$\rho_{1500}$}   & \colhead{$\rho_{912}$} & 
\colhead{}        & \colhead{$\rho_{1500}$}   & \colhead{$\rho_{912}$}}
\startdata 
DPL+DPL &
2.0$^{+0.7}_{-0.4}$\phn &
1.2$^{+0.4}_{-0.3}$\phn &
&
184$^{+6}_{-7}$\phn &
\nodata  \\
DPL+Sch &
10$^{+4}_{-3}$\phn &
5.8$^{+2.6}_{-1.8}$\phn  &
&
187$^{+6}_{-6}$\phn &
\nodata 
\enddata 
\tablecomments{All values are in units of $10^{24}$ ergs s$^{-1}$ Hz$^{-1}$ Mpc$^{-3}$.} 
\end{deluxetable} 

%% file: fig_uv_lum_den.tex
\begin{figure}[!tbp]
\includegraphics[scale=0.7,angle=0]{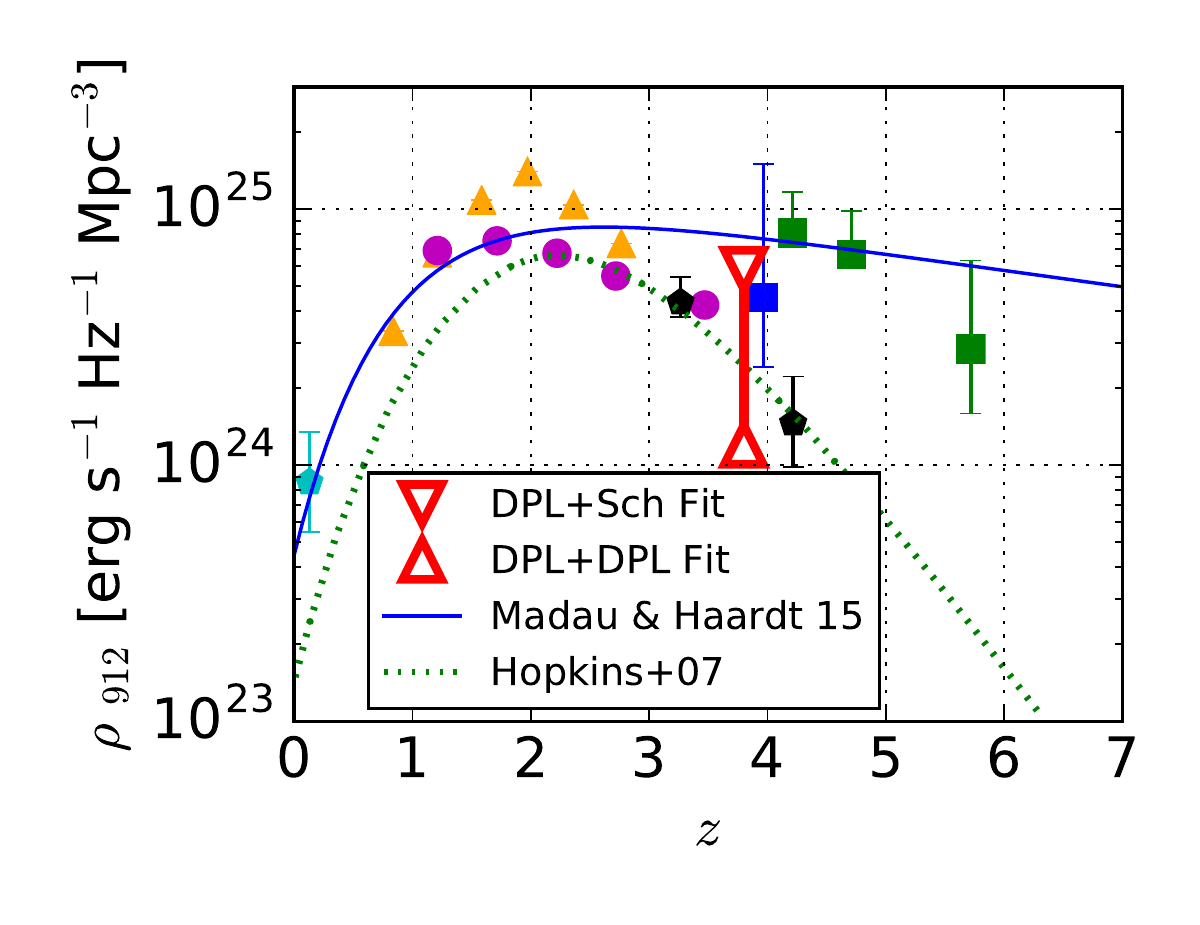}
\caption{The AGN hydrogen ionizing emissivity from this work and others. The AGN emissivity predicted by the DPL+DPL Fit and the DPL+Schechter Fit are represented as red triangles. The range they span is marked by the thick red line. The values from other works are inferred from Fig 1 of \citet*{madau15}. The original sources of each dataset are as follows: Schulze et al. (2009) (cyan pentagon), Palanque-Delabrouille et al. (2013) (orange triangles), Bongiorno et al. (2007) (magenta circles), \citet{masters12} (black pentagons), \citet{glikman11} (blue square), and \citet{giallongo15} (green squares). The solid blue line is the functional form derived by \citet*{madau15} to coincide with the plotted observation from the literature. The dotted green line shows the LyC AGN emissivity from \citet{hopkins07}. The $\rho_{912}$ from our DPL+DPL Fit is below the line by \citet{hopkins07} indicating AGNs contribute only a small fraction of the total ionizing background at $z=4$, while the $\rho_{912}$ from the DPL+Schechter fit suggests AGNs would contribute significantly to the total ionizing background.}
\label{uv_lum_den}
\end{figure}

%% file: s6_conclusions.tex
\section{Conclusions} \label{conclusions}

In this study, we measure the bright end of the rest-frame UV luminosity function of $z=4$ star-forming galaxies and the faint end of the rest-frame UV luminosity function of $z=4$ AGNs. We use nine photometric bands ($u^{\prime}g^{\prime}r^{\prime}i^{\prime}z^{\prime}$ from DECam, $J$ and $K_{s}$ from VISTA, and $3.6$ and $4.5$ $\mu$m from \textit{Spitzer}/IRAC) covering the wide area (18 deg$^2$) SHELA Field to select 3,740 candidate $z\sim4$ galaxies via a photometric redshift selection procedure. From simulations, we find a relatively low contamination rate of interloping low-$z$ galaxies and galactic stars of 20\% near our completeness limit ($m_i\sim23$) due in large part to the inclusion of IRAC photometry.

Our conclusions are as follows:
\begin{itemize}

\item We combine our candidate sample with a sample of bright AGNs from SDSS and fainter galaxies from deep Hubble imaging (including the HUDF and CANDELS) to produce a rest-frame UV luminosity function that spans the range -29$<M_{UV,i'}<$-17. This range contains both AGNs and star-forming galaxies several magnitudes above and below their respective characteristic luminosities, thus we implement a fitting procedure that simultaneously fits the AGN luminosity function and the star-forming galaxy luminosity function with independent functions. This simplifies the source selection process by not requiring a step for classifying objects as either an AGN or galaxy, which is commonly done with morphological criteria. We find the data is best fit by our DPL+DPL Fit which is a combination of a DPL function for the AGN component and a DPL function for the galaxy component. The DPL+DPL fit is preferred over the DPL+Shechter Fit, which is a combination of a DPL function for the AGN component and a Schechter function for the galaxy component, and this excess over Schechter cannot be explained by the effects of gravitational lensing. We note that we cannot significantly rule out a Schechter form.

\item We compare our measured luminosity functions to the literature and find our DPL galaxy luminosity function is in agreement with luminosity functions from the literature around the knee and to fainter magnitudes while having the shallowest bright-end slope. The AGN luminosity function from the DPL+DPL Fit has number densities similar to existing luminosity functions at magnitudes up to $M_{UV,i’}=-25.5$ mag while under predicting number densities by $\sim0.3$ dex at fainter magnitudes. The DPL+Schechter Fit has an AGN faint-end slope that is among the steepest values published. The shape of the galaxy bright end is consistent with model predictions where star formation is more efficient at higher redshift due to increased gas densities.

\item We measure $\rho_{1500}$ by integrating the rest-frame UV luminosity function fits and find that galaxies dominate the production of non-ionizing flux at $z=4$ for both possible fits. Specifically, galaxies produce a factor of $\sim$90 more non-ionizing UV output than AGNs according to the DPL+DPL Fit, while the DPL+Schechter Fit predicts galaxies produce a factor of $\sim$19 more non-ionizing UV output than AGNs.

\item We convert the AGN $\rho_{1500}$ to $\rho_{912}$ and find AGNs do not produce enough ionizing radiation to keep the universe ionized at $z=4$ by themselves if the AGN is truly represent by the DPL component in our DPL+DPL Fit. This suggests AGNs are not the dominant contributor to cosmic reionization at earlier times. On the other hand, if the DPL+Schechter Fit is true, AGNs could alone produce the $\rho_{912}$ needed to maintain the ionized state of the universe at $z=4$ and perhaps at earlier times.
\end{itemize}

Future work is needed to confirm the shape of the star-forming galaxy and AGN luminosity functions, especially where they intersect. We discuss a simple experiment to measure the relative number densities of AGNs and galaxy at luminosities where the respective luminosity functions intersect. Spectroscopic follow-up of a sample of our $z=4$ candidates in the $M_{UV,i'}\sim$-23.5 bin where our two fits predict different AGNs to galaxies ratios is underway. Imaging from space-based telescopes such as \textit{HST} or \textit{JWST} would facilitate a robust morphological classification of our bright candidates. Other possibilities for distinguishing AGNs including taking deeper X-ray imaging in the field and using \textit{JWST} to measure mid-IR SEDs \citep{kirkpatrick12}.

%% file: 1main_doc.bbl
\begin{thebibliography}{}
\expandafter\ifx\csname natexlab\endcsname\relax\def\natexlab#1{#1}\fi
\providecommand{\url}[1]{\href{#1}{#1}}
\providecommand{\dodoi}[1]{doi:~\href{http://doi.org/#1}{\nolinkurl{#1}}}
\providecommand{\doeprint}[1]{\href{http://ascl.net/#1}{\nolinkurl{http://ascl.net/#1}}}
\providecommand{\doarXiv}[1]{\href{https://arxiv.org/abs/#1}{\nolinkurl{https://arxiv.org/abs/#1}}}

\bibitem[{{Akiyama} {et~al.}(2018){Akiyama}, {He}, {Ikeda}, {Niida}, {Nagao},
  {Bosch}, {Coupon}, {Enoki}, {Imanishi}, {Kashikawa}, {Kawaguchi}, {Komiyama},
  {Lee}, {Matsuoka}, {Miyazaki}, {Nishizawa}, {Oguri}, {Ono}, {Onoue}, {Ouchi},
  {Schulze}, {Silverman}, {Tanaka}, {Tanaka}, {Terashima}, {Toba}, \&
  {Ueda}}]{akiyama17}
{Akiyama}, M., {He}, W., {Ikeda}, H., {et~al.} 2018, \pasj, 70, S34,
  \dodoi{10.1093/pasj/psx091}

\bibitem[{{Albareti} {et~al.}(2017){Albareti}, {Allende Prieto}, {Almeida},
  {Anders}, {Anderson}, {Andrews}, {Arag{\'o}n-Salamanca},
  {Argudo-Fern{\'a}ndez}, {Armengaud}, {Aubourg}, \& et~al.}]{abareti17}
{Albareti}, F.~D., {Allende Prieto}, C., {Almeida}, A., {et~al.} 2017, \apjs,
  233, 25, \dodoi{10.3847/1538-4365/aa8992}

\bibitem[{Beers {et~al.}(1990)Beers, Flynn, \& Gebhardt}]{beers90}
Beers, T.~C., Flynn, K., \& Gebhardt, K. 1990, Astronomical Journal (ISSN
  0004-6256), 100, 32

\bibitem[{{Bertin} \& {Arnouts}(1996)}]{bertin96}
{Bertin}, E., \& {Arnouts}, S. 1996, \aaps, 117, 393,
  \dodoi{10.1051/aas:1996164}

\bibitem[{{Bertin} {et~al.}(2002){Bertin}, {Mellier}, {Radovich}, {Missonnier},
  {Didelon}, \& {Morin}}]{bertin02}
{Bertin}, E., {Mellier}, Y., {Radovich}, M., {et~al.} 2002, in Astronomical
  Society of the Pacific Conference Series, Vol. 281, Astronomical Data
  Analysis Software and Systems XI, ed. D.~A. {Bohlender}, D.~{Durand}, \&
  T.~H. {Handley}, 228

\bibitem[{{Bigiel} {et~al.}(2008){Bigiel}, {Leroy}, {Walter}, {Brinks}, {de
  Blok}, {Madore}, \& {Thornley}}]{bigiel08}
{Bigiel}, F., {Leroy}, A., {Walter}, F., {et~al.} 2008, \aj, 136, 2846,
  \dodoi{10.1088/0004-6256/136/6/2846}

\bibitem[{Bouwens {et~al.}(2007)Bouwens, Illingworth, Franx, \&
  Ford}]{bouwens07}
Bouwens, R.~J., Illingworth, G.~D., Franx, M., \& Ford, H. 2007, The
  Astrophysical Journal, 670, 928

\bibitem[{{Bouwens} {et~al.}(2014){Bouwens}, {Illingworth}, {Oesch},
  {Labb{\'e}}, {van Dokkum}, {Trenti}, {Franx}, {Smit}, {Gonzalez}, \&
  {Magee}}]{bouwens14}
{Bouwens}, R.~J., {Illingworth}, G.~D., {Oesch}, P.~A., {et~al.} 2014, \apj,
  793, 115, \dodoi{10.1088/0004-637X/793/2/115}

\bibitem[{{Bouwens} {et~al.}(2015){Bouwens}, {Illingworth}, {Oesch}, {Trenti},
  {Labb{\'e}}, {Bradley}, {Carollo}, {van Dokkum}, {Gonzalez}, {Holwerda},
  {Franx}, {Spitler}, {Smit}, \& {Magee}}]{bouwens15}
---. 2015, \apj, 803, 34, \dodoi{10.1088/0004-637X/803/1/34}

\bibitem[{Boyle {et~al.}(2000)Boyle, Shanks, Croom, Smith, Miller, Loaring, \&
  Heymans}]{boyle00}
Boyle, B.~J., Shanks, T., Croom, S.~M., {et~al.} 2000, Monthly Notices of the
  Royal Astronomical Society, 317, 1014

\bibitem[{Brammer {et~al.}(2008)Brammer, van Dokkum, \& Coppi}]{brammer08}
Brammer, G.~B., van Dokkum, P.~G., \& Coppi, P. 2008, The Astrophysical
  Journal, 686, 1503

\bibitem[{Cardelli {et~al.}(1989)Cardelli, Clayton, \& Mathis}]{cardelli89}
Cardelli, J.~A., Clayton, G.~C., \& Mathis, J.~S. 1989, Astrophysical Journal,
  345, 245

\bibitem[{{Casey} {et~al.}(2014){Casey}, {Scoville}, {Sanders}, {Lee},
  {Cooray}, {Finkelstein}, {Capak}, {Conley}, {De Zotti}, {Farrah}, {Fu}, {Le
  Floc'h}, {Ilbert}, {Ivison}, \& {Takeuchi}}]{casey14}
{Casey}, C.~M., {Scoville}, N.~Z., {Sanders}, D.~B., {et~al.} 2014, \apj, 796,
  95, \dodoi{10.1088/0004-637X/796/2/95}

\bibitem[{{Conroy} \& {Gunn}(2010)}]{Conroy10}
{Conroy}, C., \& {Gunn}, J.~E. 2010, \apj, 712, 833,
  \dodoi{10.1088/0004-637X/712/2/833}

\bibitem[{{Conroy} {et~al.}(2009){Conroy}, {Gunn}, \& {White}}]{Conroy09}
{Conroy}, C., {Gunn}, J.~E., \& {White}, M. 2009, \apj, 699, 486,
  \dodoi{10.1088/0004-637X/699/1/486}

\bibitem[{Croom {et~al.}(2009)Croom, Richards, Shanks, Boyle, Strauss, Myers,
  Nichol, Pimbblet, Ross, Schneider, Sharp, \& Wake}]{croom09}
Croom, S.~M., Richards, G.~T., Shanks, T., {et~al.} 2009, Monthly Notices of
  the Royal Astronomical Society, 399, 1755

\bibitem[{Cucciati {et~al.}(2012)Cucciati, Tresse, Ilbert, Le~F{\`e}vre,
  Garilli, Le~Brun, Cassata, Franzetti, Maccagni, Scodeggio, Zucca, Zamorani,
  Bardelli, Bolzonella, Bielby, McCracken, Zanichelli, \& Vergani}]{cucciati12}
Cucciati, O., Tresse, L., Ilbert, O., {et~al.} 2012, Astronomy {\&}
  Astrophysics, 539, A31

\bibitem[{{Eisenstein} {et~al.}(2011){Eisenstein}, {Weinberg}, {Agol},
  {Aihara}, {Allende Prieto}, {Anderson}, {Arns}, {Aubourg}, {Bailey},
  {Balbinot}, \& et~al.}]{eisenstein11}
{Eisenstein}, D.~J., {Weinberg}, D.~H., {Agol}, E., {et~al.} 2011, \aj, 142,
  72, \dodoi{10.1088/0004-6256/142/3/72}

\bibitem[{{Finkelstein}(2016)}]{fink16}
{Finkelstein}, S.~L. 2016, \pasa, 33, e037, \dodoi{10.1017/pasa.2016.26}

\bibitem[{{Finkelstein} {et~al.}(2012){Finkelstein}, {Papovich}, {Salmon},
  {Finlator}, {Dickinson}, {Ferguson}, {Giavalisco}, {Koekemoer}, {Reddy},
  {Bassett}, {Conselice}, {Dunlop}, {Faber}, {Grogin}, {Hathi}, {Kocevski},
  {Lai}, {Lee}, {McLure}, {Mobasher}, \& {Newman}}]{finkelstein12}
{Finkelstein}, S.~L., {Papovich}, C., {Salmon}, B., {et~al.} 2012, \apj, 756,
  164, \dodoi{10.1088/0004-637X/756/2/164}

\bibitem[{{Finkelstein} {et~al.}(2015){Finkelstein}, {Song}, {Behroozi},
  {Somerville}, {Papovich}, {Milosavljevic}, {Dekel}, {Narayanan}, {Ashby},
  {Cooray}, {Fazio}, {Ferguson}, {Koekemoer}, {Salmon}, \& {Willner}}]{fink15}
{Finkelstein}, S.~L., {Song}, M., {Behroozi}, P., {et~al.} 2015, ArXiv
  e-prints.
\newblock \doarXiv{1504.00005}

\bibitem[{{Foreman-Mackey} {et~al.}(2013){Foreman-Mackey}, {Hogg}, {Lang}, \&
  {Goodman}}]{foreman-mackey13}
{Foreman-Mackey}, D., {Hogg}, D.~W., {Lang}, D., \& {Goodman}, J. 2013, \pasp,
  125, 306, \dodoi{10.1086/670067}

\bibitem[{{Gawiser} {et~al.}(2006){Gawiser}, {van Dokkum}, {Herrera}, {Maza},
  {Castander}, {Infante}, {Lira}, {Quadri}, {Toner}, {Treister}, {Urry},
  {Altmann}, {Assef}, {Christlein}, {Coppi}, {Dur{\'a}n}, {Franx}, {Galaz},
  {Huerta}, {Liu}, {L{\'o}pez}, {M{\'e}ndez}, {Moore}, {Rubio}, {Ruiz}, {Toft},
  \& {Yi}}]{gawiser06}
{Gawiser}, E., {van Dokkum}, P.~G., {Herrera}, D., {et~al.} 2006, \apjs, 162,
  1, \dodoi{10.1086/497644}

\bibitem[{Geach {et~al.}(2017)Geach, Lin, Makler, Kneib, Ross, Wang, Hsieh,
  Leauthaud, Bundy, McCracken, Comparat, Caminha, Hudelot, Lin, van Waerbeke,
  Pereira, \& Mast}]{geach17}
Geach, J.~E., Lin, Y.~T., Makler, M., {et~al.} 2017, The Astrophysical Journal
  Supplement Series, 231, 7

\bibitem[{Giallongo {et~al.}(2015)Giallongo, Grazian, Fiore, Fontana,
  Pentericci, Vanzella, Dickinson, Kocevski, Castellano, Cristiani, Ferguson,
  Finkelstein, Grogin, Hathi, Koekemoer, Newman, \& Salvato}]{giallongo15}
Giallongo, E., Grazian, A., Fiore, F., {et~al.} 2015, Astronomy {\&}
  Astrophysics, 578, A83

\bibitem[{Glikman {et~al.}(2011)Glikman, Djorgovski, Stern, Dey, Jannuzi, \&
  Lee}]{glikman11}
Glikman, E., Djorgovski, S.~G., Stern, D., {et~al.} 2011, The Astrophysical
  Journal Letters, 728, L26

\bibitem[{{Goodman} \& {Weare}(2010)}]{goodman10}
{Goodman}, J., \& {Weare}, J. 2010, Communications in Applied Mathematics and
  Computational Science, Vol.~5, No.~1, p.~65-80, 2010, 5, 65,
  \dodoi{10.2140/camcos.2010.5.65}

\bibitem[{{Hilbert} {et~al.}(2007){Hilbert}, {White}, {Hartlap}, \&
  {Schneider}}]{hilbert07}
{Hilbert}, S., {White}, S.~D.~M., {Hartlap}, J., \& {Schneider}, P. 2007,
  \mnras, 382, 121, \dodoi{10.1111/j.1365-2966.2007.12391.x}

\bibitem[{Hopkins {et~al.}(2007)Hopkins, Richards, \& Hernquist}]{hopkins07}
Hopkins, P.~F., Richards, G.~T., \& Hernquist, L. 2007, The Astrophysical
  Journal, 654, 731

\bibitem[{Ikeda {et~al.}(2011)Ikeda, Nagao, Matsuoka, Taniguchi, Shioya, Trump,
  Capak, Comastri, Enoki, Ideue, Kakazu, Koekemoer, Morokuma, Murayama, Saito,
  Salvato, Schinnerer, Scoville, \& Silverman}]{ikeda11}
Ikeda, H., Nagao, T., Matsuoka, K., {et~al.} 2011, The Astrophysical Journal
  Letters, 728, L25

\bibitem[{{Kennicutt} \& {Evans}(2012)}]{kennicutt12}
{Kennicutt}, R.~C., \& {Evans}, N.~J. 2012, \araa, 50, 531,
  \dodoi{10.1146/annurev-astro-081811-125610}

\bibitem[{{Kirkpatrick} {et~al.}(2012){Kirkpatrick}, {Pope}, {Alexander},
  {Charmandaris}, {Daddi}, {Dickinson}, {Elbaz}, {Gabor}, {Hwang}, {Ivison},
  {Mullaney}, {Pannella}, {Scott}, {Altieri}, {Aussel}, {Bournaud}, {Buat},
  {Coia}, {Dannerbauer}, {Dasyra}, {Kartaltepe}, {Leiton}, {Lin}, {Magdis},
  {Magnelli}, {Morrison}, {Popesso}, \& {Valtchanov}}]{kirkpatrick12}
{Kirkpatrick}, A., {Pope}, A., {Alexander}, D.~M., {et~al.} 2012, \apj, 759,
  139, \dodoi{10.1088/0004-637X/759/2/139}

\bibitem[{{Kurucz}(1979)}]{Kurucz79}
{Kurucz}, R.~L. 1979, \apjs, 40, 1, \dodoi{10.1086/190589}

\bibitem[{LaMassa {et~al.}(2016)LaMassa, Urry, Cappelluti, B{\"o}hringer,
  Comastri, Glikman, Richards, Ananna, Brusa, Cardamone, Chon, Civano, Farrah,
  Gilfanov, Green, Komossa, Lira, Makler, Marchesi, Pecoraro, Ranalli, Salvato,
  Schawinski, Stern, Treister, \& Viero}]{lamassa16}
LaMassa, S.~M., Urry, C.~M., Cappelluti, N., {et~al.} 2016, The Astrophysical
  Journal, 817, 172

\bibitem[{{Lang} {et~al.}(2016{\natexlab{a}}){Lang}, {Hogg}, \&
  {Mykytyn}}]{lang16a}
{Lang}, D., {Hogg}, D.~W., \& {Mykytyn}, D. 2016{\natexlab{a}}, {The Tractor:
  Probabilistic astronomical source detection and measurement}, Astrophysics
  Source Code Library.
\newblock \doeprint{1604.008}

\bibitem[{{Lang} {et~al.}(2016{\natexlab{b}}){Lang}, {Hogg}, \&
  {Schlegel}}]{lang16b}
{Lang}, D., {Hogg}, D.~W., \& {Schlegel}, D.~J. 2016{\natexlab{b}}, \aj, 151,
  36, \dodoi{10.3847/0004-6256/151/2/36}

\bibitem[{Le~F{\`e}vre {et~al.}(2013)Le~F{\`e}vre, Cassata, Cucciati, Garilli,
  Ilbert, Le~Brun, Maccagni, Moreau, Scodeggio, Tresse, Zamorani, Adami,
  Arnouts, Bardelli, Bolzonella, Bondi, Bongiorno, Bottini, Cappi, Charlot,
  Ciliegi, Contini, de~la Torre, Foucaud, Franzetti, Gavignaud, Guzzo, Iovino,
  Lemaux, L{\'o}pez-Sanjuan, McCracken, Marano, Marinoni, Mazure, Mellier,
  Merighi, Merluzzi, Paltani, Pello, Pollo, Pozzetti, Scaramella, Tasca,
  Vergani, Vettolani, Zanichelli, \& Zucca}]{lefevre13}
Le~F{\`e}vre, O., Cassata, P., Cucciati, O., {et~al.} 2013, Astronomy {\&}
  Astrophysics, 559, A14

\bibitem[{Lehmer {et~al.}(2012)Lehmer, Xue, Brandt, Alexander, Bauer, Brusa,
  Comastri, Gilli, Hornschemeier, Luo, Paolillo, Ptak, Shemmer, Schneider,
  Tozzi, \& Vignali}]{lehmer12}
Lehmer, B.~D., Xue, Y.~Q., Brandt, W.~N., {et~al.} 2012, The Astrophysical
  Journal, 752, 46

\bibitem[{{Leroy} {et~al.}(2008){Leroy}, {Walter}, {Brinks}, {Bigiel}, {de
  Blok}, {Madore}, \& {Thornley}}]{leroy08}
{Leroy}, A.~K., {Walter}, F., {Brinks}, E., {et~al.} 2008, \aj, 136, 2782,
  \dodoi{10.1088/0004-6256/136/6/2782}

\bibitem[{{Liddle}(2007)}]{liddle07}
{Liddle}, A.~R. 2007, \mnras, 377, L74,
  \dodoi{10.1111/j.1745-3933.2007.00306.x}

\bibitem[{{Lima} {et~al.}(2010){Lima}, {Jain}, \& {Devlin}}]{lima10}
{Lima}, M., {Jain}, B., \& {Devlin}, M. 2010, \mnras, 406, 2352,
  \dodoi{10.1111/j.1365-2966.2010.16884.x}

\bibitem[{{Madau}(1995)}]{madau95}
{Madau}, P. 1995, \apj, 441, 18, \dodoi{10.1086/175332}

\bibitem[{Madau \& Haardt(2015)}]{madau15}
Madau, P., \& Haardt, F. 2015, The Astrophysical Journal Letters, 813, L8

\bibitem[{{Marchesini} {et~al.}(2009){Marchesini}, {van Dokkum}, {F{\"o}rster
  Schreiber}, {Franx}, {Labb{\'e}}, \& {Wuyts}}]{marchesini09}
{Marchesini}, D., {van Dokkum}, P.~G., {F{\"o}rster Schreiber}, N.~M., {et~al.}
  2009, \apj, 701, 1765, \dodoi{10.1088/0004-637X/701/2/1765}

\bibitem[{Masters {et~al.}(2012)Masters, Capak, Salvato, Civano, Mobasher,
  Siana, Hasinger, Impey, Nagao, Trump, Ikeda, Elvis, \& Scoville}]{masters12}
Masters, D., Capak, P., Salvato, M., {et~al.} 2012, The Astrophysical Journal,
  755, 169

\bibitem[{McGreer {et~al.}(2017)McGreer, Fan, Jiang, \& Cai}]{mcgreer17}
McGreer, I.~D., Fan, X., Jiang, L., \& Cai, Z. 2017, arXiv.org

\bibitem[{{Muzzin} {et~al.}(2013){Muzzin}, {Marchesini}, {Stefanon}, {Franx},
  {McCracken}, {Milvang-Jensen}, {Dunlop}, {Fynbo}, {Brammer}, {Labb{\'e}}, \&
  {van Dokkum}}]{muzzin13}
{Muzzin}, A., {Marchesini}, D., {Stefanon}, M., {et~al.} 2013, \apj, 777, 18,
  \dodoi{10.1088/0004-637X/777/1/18}

\bibitem[{Niida {et~al.}(2016)Niida, Nagao, Ikeda, Matsuoka, Kobayashi, Toba,
  \& Taniguchi}]{niida16}
Niida, M., Nagao, T., Ikeda, H., {et~al.} 2016, The Astrophysical Journal, 832,
  208

\bibitem[{{Oke} \& {Gunn}(1983)}]{oke83}
{Oke}, J.~B., \& {Gunn}, J.~E. 1983, \apj, 266, 713, \dodoi{10.1086/160817}

\bibitem[{{Ono} {et~al.}(2018){Ono}, {Ouchi}, {Harikane}, {Toshikawa}, {Rauch},
  {Yuma}, {Sawicki}, {Shibuya}, {Shimasaku}, {Oguri}, {Willott}, {Akhlaghi},
  {Akiyama}, {Coupon}, {Kashikawa}, {Komiyama}, {Konno}, {Lin}, {Matsuoka},
  {Miyazaki}, {Nagao}, {Nakajima}, {Silverman}, {Tanaka}, {Taniguchi}, \&
  {Wang}}]{ono17}
{Ono}, Y., {Ouchi}, M., {Harikane}, Y., {et~al.} 2018, \pasj, 70, S10,
  \dodoi{10.1093/pasj/psx103}

\bibitem[{{Papovich} {et~al.}(2016){Papovich}, {Shipley}, {Mehrtens}, {Lanham},
  {Lacy}, {Ciardullo}, {Finkelstein}, {Bassett}, {Behroozi}, {Blanc}, {de
  Jong}, {DePoy}, {Drory}, {Gawiser}, {Gebhardt}, {Gronwall}, {Hill}, {Hopp},
  {Jogee}, {Kawinwanichakij}, {Marshall}, {McLinden}, {Mentuch Cooper},
  {Somerville}, {Steinmetz}, {Tran}, {Tuttle}, {Viero}, {Wechsler}, \&
  {Zeimann}}]{papovich16}
{Papovich}, C., {Shipley}, H.~V., {Mehrtens}, N., {et~al.} 2016, \apjs, 224,
  28, \dodoi{10.3847/0067-0049/224/2/28}

\bibitem[{Pedregosa {et~al.}(2011)Pedregosa, Varoquaux, Gramfort, Michel,
  Thirion, Grisel, Blondel, Prettenhofer, Weiss, Dubourg, Vanderplas, Passos,
  Cournapeau, Brucher, Perrot, \& Duchesnay}]{scikit-learn}
Pedregosa, F., Varoquaux, G., Gramfort, A., {et~al.} 2011, Journal of Machine
  Learning Research, 12, 2825

\bibitem[{{Planck Collaboration} {et~al.}(2014){Planck Collaboration}, {Ade},
  {Aghanim}, {Armitage-Caplan}, {Arnaud}, {Ashdown}, {Atrio-Barandela},
  {Aumont}, {Baccigalupi}, {Banday}, \& et~al.}]{planck13}
{Planck Collaboration}, {Ade}, P.~A.~R., {Aghanim}, N., {et~al.} 2014, \aap,
  571, A16, \dodoi{10.1051/0004-6361/201321591}

\bibitem[{Quadri {et~al.}(2006)Quadri, Marchesini, van Dokkum, Gawiser, Franx,
  Lira, Rudnick, Urry, Maza, Kriek, Barrientos, Blanc, Castander, Christlein,
  Coppi, Hall, Herrera, Infante, Taylor, Treister, \& Willis}]{quadri06}
Quadri, R., Marchesini, D., van Dokkum, P., {et~al.} 2006, arXiv.org, 1103

\bibitem[{Richards {et~al.}(2006)Richards, Strauss, Fan, Hall, Jester,
  Schneider, Berk, Stoughton, Anderson, Brunner, Gray, Gunn, Ivezi{\'c},
  Kirkland, Knapp, Loveday, Meiksin, Pope, Szalay, Thakar, Yanny, York, \&
  Collaboration}]{richards06}
Richards, G.~T., Strauss, M.~A., Fan, X., {et~al.} 2006, arXiv.org, 2766

\bibitem[{Schechter(1976)}]{schechter76}
Schechter, P. 1976, Astrophysical Journal, 203, 297

\bibitem[{Schlafly \& Finkbeiner(2011)}]{schlafly11}
Schlafly, E.~F., \& Finkbeiner, D.~P. 2011, The Astrophysical Journal, 737, 103

\bibitem[{Schneider {et~al.}(2010)Schneider, Richards, Hall, Strauss, Anderson,
  Boroson, Ross, Shen, Brandt, Fan, Inada, Jester, Knapp, Krawczyk, Thakar,
  Vanden~Berk, Voges, Yanny, York, Bahcall, Bizyaev, Blanton, Brewington,
  Brinkmann, Eisenstein, Frieman, Fukugita, Gray, Gunn, Hibon, Ivezi{\'c},
  Kent, Kron, Lee, Lupton, Malanushenko, Malanushenko, Oravetz, Pan, Pier,
  Price, Saxe, Schlegel, Simmons, Snedden, SubbaRao, Szalay, \&
  Weinberg}]{schneider10}
Schneider, D.~P., Richards, G.~T., Hall, P.~B., {et~al.} 2010, The Astronomical
  Journal, 139, 2360

\bibitem[{Shull {et~al.}(2012)Shull, Stevans, \& Danforth}]{Shull:2012}
Shull, J.~M., Stevans, M., \& Danforth, C.~W. 2012, The Astrophysical Journal,
  752, 162

\bibitem[{{Smith} {et~al.}(2016){Smith}, {Windhorst}, {Jansen}, {Cohen},
  {Jiang}, {Dijkstra}, {Koekemoer}, {Bielby}, {Inoue}, {MacKenty}, {O'Connell},
  \& {Silk}}]{smith16}
{Smith}, B.~M., {Windhorst}, R.~A., {Jansen}, R.~A., {et~al.} 2016, ArXiv
  e-prints.
\newblock \doarXiv{1602.01555}

\bibitem[{{Somerville} {et~al.}(2015){Somerville}, {Popping}, \&
  {Trager}}]{somerville15}
{Somerville}, R.~S., {Popping}, G., \& {Trager}, S.~C. 2015, \mnras, 453, 4337,
  \dodoi{10.1093/mnras/stv1877}

\bibitem[{{Song} {et~al.}(2016){Song}, {Finkelstein}, {Ashby}, {Grazian}, {Lu},
  {Papovich}, {Salmon}, {Somerville}, {Dickinson}, {Duncan}, {Faber}, {Fazio},
  {Ferguson}, {Fontana}, {Guo}, {Hathi}, {Lee}, {Merlin}, \&
  {Willner}}]{song16}
{Song}, M., {Finkelstein}, S.~L., {Ashby}, M.~L.~N., {et~al.} 2016, \apj, 825,
  5, \dodoi{10.3847/0004-637X/825/1/5}

\bibitem[{{Steidel} {et~al.}(1999){Steidel}, {Adelberger}, {Giavalisco},
  {Dickinson}, \& {Pettini}}]{steidel99}
{Steidel}, C.~C., {Adelberger}, K.~L., {Giavalisco}, M., {Dickinson}, M., \&
  {Pettini}, M. 1999, \apj, 519, 1, \dodoi{10.1086/307363}

\bibitem[{{Steidel} {et~al.}(1996){Steidel}, {Giavalisco}, {Dickinson}, \&
  {Adelberger}}]{steidel96}
{Steidel}, C.~C., {Giavalisco}, M., {Dickinson}, M., \& {Adelberger}, K.~L.
  1996, \aj, 112, 352, \dodoi{10.1086/118019}

\bibitem[{Stevans {et~al.}(2014)Stevans, Shull, Danforth, \&
  Tilton}]{stevans:2014}
Stevans, M.~L., Shull, J.~M., Danforth, C.~W., \& Tilton, E.~M. 2014, The
  Astrophysical Journal, 794, 75

\bibitem[{Szalay {et~al.}(1999)Szalay, Connolly, \& Szokoly}]{szalay99}
Szalay, A.~S., Connolly, A.~J., \& Szokoly, G.~P. 1999, The Astronomical
  Journal, 117, 68

\bibitem[{Telfer {et~al.}(2002)Telfer, Zheng, Kriss, \& Davidsen}]{Telfer:2002}
Telfer, R.~C., Zheng, W., Kriss, G.~A., \& Davidsen, A.~F. 2002, The
  Astrophysical Journal, 565, 773

\bibitem[{{van der Burg} {et~al.}(2010){van der Burg}, {Hildebrandt}, \&
  {Erben}}]{van10}
{van der Burg}, R.~F.~J., {Hildebrandt}, H., \& {Erben}, T. 2010, \aap, 523,
  A74, \dodoi{10.1051/0004-6361/200913812}

\bibitem[{Vanden~Berk {et~al.}(2001)Vanden~Berk, Richards, Bauer, Strauss,
  Schneider, Heckman, York, Hall, Fan, Knapp, Anderson, Annis, Bahcall,
  Bernardi, Briggs, Brinkmann, Brunner, Burles, Carey, Castander, Connolly,
  Crocker, Csabai, Doi, Finkbeiner, Friedman, Frieman, Fukugita, Gunn,
  Hennessy, Ivezi{\'c}, Kent, Kunszt, Lamb, Leger, Long, Loveday, Lupton,
  Meiksin, Merelli, Munn, Newberg, Newcomb, Nichol, Owen, Pier, Pope, Rockosi,
  Schlegel, Siegmund, Smee, Snir, Stoughton, Stubbs, SubbaRao, Szalay, Szokoly,
  Tremonti, Uomoto, Waddell, Yanny, \& Zheng}]{VandenBerk:2001}
Vanden~Berk, D.~E., Richards, G.~T., Bauer, A., {et~al.} 2001, The Astronomical
  Journal, 122, 549

\bibitem[{Viironen {et~al.}(2017)Viironen, L{\'o}pez-Sanjuan,
  Hern{\'a}ndez-Monteagudo, Chaves-Montero, Ascaso, Bonoli,
  Crist{\'o}bal-Hornillos, D{\'\i}az-Garc{\'\i}a, Fern{\'a}ndez-Soto,
  M{\'a}rquez, Masegosa, Povi{\'c}, Varela, Cenarro, Aguerri, Alfaro,
  Aparicio-Villegas, Ben{\'\i}tez, Broadhurst, Cabrera-Ca{\~n}o, Castander,
  Cepa, Cervi{\~n}o, Gonz{\'a}lez~Delgado, Husillos, Infante, Mart{\'\i}nez,
  Moles, Molino, Del~Olmo, Perea, Prada, \& Quintana}]{viironen17}
Viironen, K., L{\'o}pez-Sanjuan, C., Hern{\'a}ndez-Monteagudo, C., {et~al.}
  2017, arXiv.org, arXiv:1712.01028

\bibitem[{{White} {et~al.}(2015){White}, {Somerville}, \& {Ferguson}}]{white15}
{White}, C.~E., {Somerville}, R.~S., \& {Ferguson}, H.~C. 2015, \apj, 799, 201,
  \dodoi{10.1088/0004-637X/799/2/201}

\bibitem[{{Wyithe} {et~al.}(2011){Wyithe}, {Yan}, {Windhorst}, \&
  {Mao}}]{wyithe11}
{Wyithe}, J.~S.~B., {Yan}, H., {Windhorst}, R.~A., \& {Mao}, S. 2011, \nat,
  469, 181, \dodoi{10.1038/nature09619}

\bibitem[{{Yung} {et~al.}(2018){Yung}, {Somerville}, {Finkelstein}, {Popping},
  \& {Dav{\'e}}}]{yung18}
{Yung}, L.~Y.~A., {Somerville}, R.~S., {Finkelstein}, S.~L., {Popping}, G., \&
  {Dav{\'e}}, R. 2018, ArXiv e-prints.
\newblock \doarXiv{1803.09761}

\bibitem[{{Zacharias} {et~al.}(2004){Zacharias}, {Monet}, {Levine}, {Urban},
  {Gaume}, \& {Wycoff}}]{Zacharias04}
{Zacharias}, N., {Monet}, D.~G., {Levine}, S.~E., {et~al.} 2004, in Bulletin of
  the American Astronomical Society, Vol.~36, American Astronomical Society
  Meeting Abstracts, 1418

\end{thebibliography}
